\documentclass{aa}  
\usepackage{graphicx}
\usepackage{txfonts}
\usepackage{multirow}
\usepackage{booktabs}
\usepackage[utf8]{inputenc}
\usepackage{newunicodechar} 
\usepackage{amsmath}
\newunicodechar{⌀}{\ensuremath{\diameter}}
\DeclareUnicodeCharacter{00A0}{~}	
\usepackage{natbib}
\bibpunct{(}{)}{;}{a}{}{,} 
\usepackage[colorlinks=true, linkcolor = blue, urlcolor=blue, citecolor = blue]{hyperref}
\begin{document} 
	
	\title{In pursuit of giants:}
	
	\subtitle{
	II.	Evolution of dusty quiescent galaxies over the last six billion years from the hCOSMOS survey 
	}

	\author{D. Donevski\inst{1,2,3}
	\and I. Damjanov\inst{4,5}
	\and A. Nanni\inst{1}
	\and A. Man\inst{6}
	\and M. Giulietti\inst {2,7}
	\and M. Romano\inst{1,8}
	\and A. Lapi\inst{2,3}
	\and D. Narayanan\inst{9,10}
	\and R. Dav\'e\inst{11,12}
    \and I. Shivaei\inst{13}
	\and J. Sohn \inst{14,15}
	\and Junais\inst{1}
	\and L. Pantoni\inst{16}	
	\and Q. Li\inst{17}
}

\institute{National Centre for Nuclear Research, Pasteura 7, 02-093 Warsaw, Poland \\
	\email{darko.donevski@ncbj.gov.pl}
	\and
	SISSA, Via Bonomea 265, 34136 Trieste, Italy
	\and
	IFPU - Institute for fundamental physics of the Universe, Via Beirut 2,
	34014 Trieste, Italy
	\and
	Department of Astronomy and Physics, Saint Mary’s University, 923 Robie Street, Halifax, NS B3H 3C3, Canada
	\and
	Canada Research Chair in Astronomy and Astrophysics, Tier II
	\and
	Department of Physics \& Astronomy, University of British Columbia, 6224 Agricultural Road, Vancouver, BC V6T 1Z1, Canada
	\and
	INAF - Osservatorio di Astrofisica e Scienza dello Spazio, Via Gobetti 93/3, I-40129, Bologna, Italy
	\and
	INAF, OAPD, Vicolo dell'Osservatorio, 5, 35122 Padova, Italy
	\and
	Department of Astronomy, University of Florida, 211 Bryant Space Sciences Center, Gainesville, FL, USA
	\and
	Cosmic Dawn Center (DAWN), Copenhagen, Denmark
	\and 
	Institute for Astronomy, Royal Observatory, University of Edinburgh,
	Edinburgh EH9 3HJ, UK
	\and
	Department of Physics \& Astronomy, University of the Western Cape,
	Robert Sobukwe Rd, Bellville, 7535, South Africa
	\and
	Steward Observatory, University of Arizona, Tucson, AZ 85721, USA
	\and
	Astronomy Program, Department of Physics and Astronomy, Seoul National University, 1 Gwanak-ro, Gwanak-gu, Seoul 08826, Republic of Korea
	\and 
	Smithsonian Astrophysical Observatory, 60 Garden Street, Cambridge, MA 02138, USA
	\and
	CEA, CNRS, Universit\'e Paris-Saclay, 91191, Gif-sur-Yvette, France
	\and
	Max-Planck-Institute f\"ur Astrophysik, Karl-Schwarzschild-Strasse 1, D-85740 Garching, Germany
}

	\date{Received ; accepted }
	
	
	\abstract
	{
		 The physical mechanisms that link the termination of star formation in quiescent galaxies and the evolution of their baryonic components, stars and the interstellar medium (ISM; dust, gas and metals), are poorly constrained beyond the local universe. In this work, we characterise the evolution of the dust content in 548 quiescent galaxies observed at $0.1<z<0.6$ as part of the hCOSMOS spectroscopic redshift survey. This is the largest sample of quiescent galaxies at intermediate redshifts, for which the dust, stellar and metal abundances are consistently estimated to date. We analyse how the crucial markers of a galaxy dust life-cycle, such as specific dust mass ($M_{\rm dust}$/$M_{\rm \star}$), evolve with different physical parameters, i.e., gas-phase metallicity ($Z_{\rm gas}$), time since quenching ($t_{\rm quench}$), stellar mass ($M_{\star}$) and stellar population age.
		
		
		
		We find morphology to be important factor of a large scatter ($\sim2$ orders of magnitude) in $M_{\rm dust}/M_{\rm \star}$. Quiescent spirals exhibit strong evolutionary trends of specific dust mass with $M_{\star}$, stellar age and galaxy size, in contrast to little to no evolution in ellipticals. When transitioning from solar to super-solar metallicities ($8.7\lesssim \rm12+\log(O/H)\lesssim9.1$), quiescent spirals undergo reversal of $M_{\rm dust}/M_{\rm \star}$, indicative of a change in dust production efficiency. Through modelling the star formation histories of our objects we unveil a broad dynamical range of post-quenching timescales ($60\:\rm Myr<t_{\rm quench}<3.2\:\rm Gyr$). We show that $M_{\rm dust}/M_{\rm \star}$ is the highest in recently quenched systems ($t_{\rm quench}<500$ Myr), but its further evolution is non-monotonic as a consequence of different pathways for dust formation, growth, or removal on various timescales. 
		
		
		Our data are best described by simulations that include dust growth in the ISM. While this process is prevalent in the majority of galaxies, for $\sim15\%$ of objects we find evidence of additional dust content acquired externally, most likely via minor mergers. Altogether, our results strongly suggest that prolonged dust production on a timescale $0.5-1\:\rm Gyr$ since quenching may be common in dusty quiescent galaxies at intermediate redshifts, even if their gas reservoirs are heavily exhausted (i.e., cold gas fraction $<1-5\%$). 
	}

	
	\maketitle
	
	\section{Introduction}
	
	The general population of passively evolved galaxies (the so-called quiescent galaxies, QGs) are known to have no or little ongoing star-formation activity, which usually places them below the star-forming main-sequence (MS; e.g., \citealt{daddi05}, \citealt{toft07}, \citealt{dokkum10}, \citealt{schreiber15}). Whatever physical mechanism is in place, one of the important outcomes of quenching star-formation is the evolution of its agents - gas and dust in the cold interstellar medium (ISM). 
	
	The abundance of dust in the ISM is regulated by a complex network of dust formation and destruction channels (see e.g., \citealt{zhukovska16} and references therein). Even though many details of the involved physical and chemical processes are not fully understood, the study of dust in galaxies is essential to understand their evolution for a number of reasons. On one hand, dust is critical in the thermal balance of gas and in shielding the dense clouds from ultraviolet (UV) radiation, supporting the star-formation (\citealt{cuppen17}). Dust further affects the spectral energy distributions (SED) of massive galaxies to the extent that at shorter wavelengths, stellar light is absorbed by dust and re-emitted in the far-infrared (FIR). On the other hand, the interplay of dust with metals and gas in the ISM constitutes a vital component of the baryon cycle. Metals in the gas phase are proposed to play a role in the growth of dust particles (\citealt{asano13}, \citealt{slavin15}, \citealt{zhukovska16}, \citealt{hirashita17}, \citealt{popping17}, \citealt{pantoni19}), and dust itself is an important constituent of the cold ISM, often considered its reliable tracer not only in star-forming galaxies (e.g., \citealt{scoville17}), but also in QGs (\citealt{magdis21}). 
	
	Recent advent of IR/sub-mm instruments such as \textit{Herschel}, NOEMA and ALMA allowed us to identify QGs with substantial masses of dust and molecular gas ($M_{\rm dust}>10^{7}, M_{\odot}$, $M_{\rm gas}>10^{9} M_{\odot}$, \citealt{smith14}, \citealt{gobat18}, \citealt{belli21}). These findings challenged the historical picture of QGs containing very little ISM material in proportion to their stellar mass. The vast majority of known studies have focused on studying the dust and gas in the QGs in the local Universe ($z\sim0$) either through galaxy FIR continuum emission (e.g., \citealt{smith14}, \citealt{rowlands15}, \citealt{michalowski19}), and/or via low transition CO lines (e.g. \citealt{davis13}, \citealt{deVis17}, \citealt{lipsb19}, \citealt{sansom19}). Moreover, several recent works use extensive ALMA and NOEMA follow-ups to identify small number of dusty QGs in the distant Universe ($z>1-3$, \citealt{williams21}, \citealt{whitaker21b}). Nevertheless, there is no consensus view on whether the observed dust in QGs is solely of internal origin (due to past star-formation), or whether external contributions (i.e., from mergers) are influential as well. It is also debatable how long the dust and gas in the ISM will survive after the quenching. Some studies report significant dust and gas content and prolonged timescales for ISM removal ($\gtrsim 1$ Gyr; \citealt{rowlands15},  \citealt{rudnick17}, \citealt{gobat18}, \citealt{michalowski19}, \citealt{woodrum22}), whereas other report rapid depletion of dust and gas ($\lesssim 100-500$ Myr;  \citealt{sargent15}, \citealt{williams21}, \citealt{whitaker21b}). 
	
	Quantifying changes in dust abundance as galaxies evolve is of particular interest when studying dusty QGs. To that end, specific dust masses ($M_{\rm dust}/M_{\star}$) have been proposed as a useful marker of dust life-cycle against multiple destruction processes (\citealt{calura17}, \citealt{donevski20}). \cite{michalowski19} use the anti-correlation of $M_{\rm dust}/M_{\star}$ with stellar age to derive the time for dust removal in QGs and to understand its connection with the shutdown of star-formation. When $M_{\rm gas}$ is unavailable, the $M_{\rm dust}/M_{\star}$ is often used as an observational proxy for molecular gas fraction (\citealt{remyruyer14}, \citealt{magdis21}). Furthermore, as dust depletes metals from the gas-phase ISM, adding the information of gas metallicity ($Z_{\rm gas}$) to the relation between $M_{\rm dust}$ and $M_{\star}$ allows us to better constrain the dominant dust formation mechanisms (\citealt{feldman15}, \citealt{deVis17}). Beyond the local universe, such connection was only studied in the dusty star-forming galaxies (DSFGs) observed with ALMA at $z\sim2.3$ (\citealt{shivaei22}). Despite its importance, the evolution of specific dust masses as a function of both redshift and physical quantities (in particular, $Z_{\rm gas}$, stellar age and time since quenching) has not yet been constrained for a statistical sample of QGs at intermediate redshifts. The task has multiple challenges: (1) To properly determine dust quantities (i.e., dust luminosity ($L_{\rm IR}$) and $M_{\rm dust}$) well-sampled IR SEDs towards the Rayleigh-Jeans (RJ) tail are required; (2) QGs have considerably fainter IR continuum emission than DSFGs, and careful de-blending of fluxes obtained with single-dish instruments is required to model the IR SEDs (e.g., \citealt{man16}, \citealt{galliano21}); (3) It has been demonstrated that obtaining direct gas-metallicity measurements for QGs is a demanding task (\citealt{metalsqg19}, \citealt{kumari21}). To partially overcome those issues, \cite{magdis21} applied multi-wavelength stacking analysis on massive ($\log(M_{\star}/M_{\odot})>10.8$) QGs from the COSMOS field up to $z\sim1.5$. They reported a sharp decline of $M_{\rm dust}/M_{\star}$ from $z\sim1$ to $z=0$, interpreting it as a consequence of rapid exhaustion of molecular gas. However, stacking does not allow the investigation of physical markers of dust evolution in individual sources. This prevents us from answering important questions, such as: \textit{What processes of dust production/removal are dominant in QGs during their evolution?}
	
	From the theoretical standpoint, there is a growing number of studies that have modelled the dust evolution in different classes of cosmological simulations (\citealt{mckinon17}, \citealt{simba},  \citealt{hou19}, \citealt{graziani19}, \citealt{aoyama19}), semi-analytic models (\citealt{lacey16}, \citealt{popping17}, \citealt{cousin19}, \citealt{vijayan19}, \citealt{lagos19}, \citealt{pantoni19}) and chemical evolution models (\citealt{asano13}, \citealt{nanni20}). While some of them show success in reproducing dust related properties of star-forming galaxies, they still lack observational constraints from statistical samples of QGs at $z>0$. On top of this, there is an active debate regarding the dominant dust producers in massive galaxies. One group of models advocates for major contribution from stellar sources (i.e., either from asymptotic giant branch (AGB) stars (e.g., \citealt{clemens10}, \citealt{valiante11}) or supernovae (SNe; \citealt{gall18})), while others predict prevalence of the dust grain growth in the ISM (e.g., \citealt{asano13}, \citealt{hirashita15}). The latter channel is thought to be important at late cosmic epochs ($z<2$), but it is strongly dependent on "critical" gas-metallicity beyond which it takes over stellar production (\citealt{triani20}). Thus, the observed co-evolution between the $M_{\rm dust}, M_{\star}$ and $Z_{\rm gas}$ in QGs is crucial to challenge (and eventually rule out) some of the prescriptions related to the growth and destruction of dust.
	
	The present paper addresses these challenges. The first major question we aim to address is how a specific dust mass evolves since quenching in individually detected QGs. We assembled a statistical data set of spectroscopically selected dusty QGs at $0.1<z<0.6$ to achieve this goal. We complement deep multi-wavelength catalogue with the IR fluxes of the carefully de-blended sources and apply physically motivated SED modelling to self-consistently derive sources' physical properties. By considering independent spectroscopic information such as $Z_{\rm gas}$ and stellar age index ($D_{\rm n}4000$), we study how the $M_{\rm dust}/M_{\rm \star}$ changes with metallicity, stellar mass and time since quenching. We then address the second big question, namely, how the observed evolution of dust properties in QGs can be understood within the framework of galaxy formation and evolution. 
	
	The paper is organised as follows. In \hyperref[sec:2]{Section 2,} we describe the data analysed in this work. In \hyperref[sec:3]{Section 3,} we explain the SED fitting methodology and provide statistical properties for our sample. In \hyperref[sec:4]{Section 4,} we present how the $M_{\rm dust}/M_{\rm \star}$ in the general population of dusty QGs scale with the galaxy redshift and sSFR. In \hyperref[sec:5]{Section 5,} we investigate the morphological impact on the evolution of the $M_{\rm dust}/M_{\rm \star}$ with physical parameters such as $Z_{\rm gas}$ and $M_{\star}$. The interpretation of results with state-of-the-art simulations is presented in \hyperref[sec:6]{Section 6,}, while our main conclusions are outlined in \hyperref[sec:7]{Section 7}. Throughout the paper, we assume a \cite{planck16} cosmology and the Chabrier initial mass function (IMF; \citealt{chabrier03}). 
	
	\section{Data and sample selection}
	\label{sec:2}
	
	The starting point of our sample is the hCOSMOS spectroscopic survey catalogue (for comprehensive description of the survey, target selection and observational design, see \citealt{damjanov18}). Briefly, the hCOSMOS is a dense spectroscopic survey, designed for providing spectra of galaxies from the COSMOS UltraVISTA catalogue (\citealt{muzzin13}). The hCOSMOS survey is the magnitude limited survey with no colour selection, and targets UltraVISTA galaxies that have $r-$band magnitude within the range of $17.8<r<21.3$. The bright limit of $r = 17.8$ is set to match the limiting magnitude of the Sloan Digital Sky Survey (SDSS) main galaxy sample. 
	The hCOSMOS survey was done with the multi-fiber fed spectrograph Hectospec, mounted on the 6.5 m MMT (\citealt{fabricant05}). Hectospec is a 300-fiber optical spectrograph with an $\sim1\rm\:deg^{2}$ field of view (FoV) and a fiber diameter of $1.5^{\prime\prime}$. In total, hCOSMOS survey observed $\sim 1.5\rm\:deg^{2}$ covering the wavelength range between 370 nm and 910 nm at a resolution of R$\sim1500$. The hCOSMOS catalogue includes 4362 galaxies with a science quality spectra. The galaxies are identified across the redshift range of $0.01<z<0.7$.
	
We supplement each source from the hCOSMOS sample with rich panchromatic photometry from publicly available photometric catalogues. Namely, for SED modelling of the data we adopt homogeneously calibrated multi-wavelength catalogues released by the Herschel Extragalactic Legacy Project (\citealt{malek19}, \citealt{shirley19}). The catalogue is created with use of the state-of-the-art de-blending method and cross-matching procedure of individual galaxies across broad wavelengths. In particular, the de-blending of confusion limited PACS (100, 160$\mu$m), and SPIRE (250, 350, and 500 $\mu$m) maps is performed with the probabilistic de-blending code \texttt{XID+} (\citealt{xid+17}). The code exploits prior information on the redshift and source position available from the higher-resolution maps observed at 3.6 $\mu$m, and extracts PACS and SPIRE fluxes beyond the conventional $\mathit{Herschel}$ confusion limit, achieving $1\sigma$ cut depth of $\sim1$ mJy at 250 $\mu$m. The detailed de-blending procedure are explained in \cite{pearson18} and \cite{shirley19}, who demonstrate that such approach takes the big advantage of fluctuations within the confused maps and place strong constraints on the peak of sources' FIR SEDs even in IR-faint galaxies. All hCOSMOS sources have wealthy multiwavelength coverage that consists of at least 15 photometric bands: along with their information in the IR part of the spectra, the optical to mid-IR data come from Subaru Suprime-Cam (six broadband filters, $V,g,r,i,z$ and $Y$), VISTA ($J, H, K_{s}$ bands), $\textit{Spitzer}$ IRAC (four bands) and $\textit{Spitzer}$ MIPS. 

We define the sample of QGs on the basis of a spectral indicator $D_{\rm n}4000$, which is a ratio of flux in the 4000-4100 $\AA$ and 3850-3950 $\AA$ bands. The $D_{\rm n}4000$ measures the strength of the 4000 $\AA$  break produced by a large number of absorption lines where ionised metals are the main contributors to the opacity. Because the strength of the 4000 $\AA$ break increases with the population age, it is always used as an indicator of a quiescence. It is worth noting that $D_{\rm n}4000$ is a strong evolutionary marker because it is insensitive to reddening and does not require a K-correction, opposite to galaxy colours. From the full hCOSMOS catalogue of 4362 sources, we select QGs as those with $D_{\rm n}4000>1.5$ (1737 sources in total). This criterion is commonly used for selecting QGs. The choice of using $D_{\rm n}4000>1.5$ to select QGs is motivated by findings from large spectroscopic surveys demonstrating that $D_{\rm n}4000$ is strongly bimodal, and offers a robust division between emission line SF galaxies and QGs at $D_{\rm n}4000\sim1.5$ (e.g., \citealt{zahid16}, \citealt{haines16}, \citealt{damjanov18}, \citealt{wu18}, \citealt{utsumi21}, \citealt{wu21}). Additionally, it has been shown that selections based on broadband colours alone are sometimes inefficient in identifying the recently quenched galaxies (\citealt{vergani18}, \citealt{wu20}). The selection based on $D_{\rm n}4000$ is also proven to be very robust and effective when combined with the equivalent width $\rm H{\delta}$ absorption line (EW($\rm H \delta$)). On average, the S/N of individual hCOSMOS spectra is not high enough for measuring EW($\rm H \delta$) for the full parent mass-complete sample, but we note that recent deep spectroscopy studies of QGs at $z\sim0.8$ from the LEGA-C survey unveil the tight relation between EW($\rm H \delta$) and $D_{\rm n}4000$ at $D_{\rm n}4000\gtrsim1.5$ (\citealt{wu18}, \citealt{zhang22}). Nevertheless, all QG selection criteria produce samples with some levels of contamination from SF outliers. For the full hCOSMOS catalogue, \cite{damjanov18} compared the identification of QGs and SF galaxies on the basis of their rest-frame UVJ colours with the identification based on $D_{\rm n}4000$, and found only $\sim13\%$ of contaminants (galaxies that have $D_{\rm n}4000>1.5$, but their UVJ colours are consistent with being SF). Interestingly, the reported fraction is lower than for any of the colour-based classifiers tested in \cite{moresco13}, who applied a variety of methods to select QGs up to $z\sim 0.5$. 
	
Finally, from all selected QGs, we build the sample of \textit{"dusty QGs"} suitable for our analysis, by imposing source detections in the IR with $\rm S/N\geq3$ in at least four photometric bands in the mid-IR-to-FIR range ($8\:\mu$m < $\lambda$ <$500\:\mu$m).\footnote[2]{When available, for the purpose of our SED fitting we also included upper limits based on $850\mu$m non-detections from the SCUBA-2 Cosmology Legacy Survey (S2CLS, \citealt{geach17}).}. These requirements are important for achieving the robustness to dust-related physical parameters estimated from SED fitting, in particular $M_{\rm dust}$ (see e.g., \citealt{berta16}). From the parent sample of 1737 QGs, we find detectable dust in 618 galaxies ($35\%$ of total). Because our goal is to explore the interplay between the stars, dust and metals in individually detected objects, we further chose the sources with reliable measurements of gas-phase metallicity ($Z_{\rm gas}$, expressed as $12+\log(\rm O/H)$). Gas phase metallicities of hCOSMOS galaxies are derived by following the method described in \citet{Zahid13} and \citet{Sohn19}. The main details of the procedure are introduced in \hyperref[App.A]{Appendix A}. In this way, we slightly narrowed the list to 603 dusty QGs for the further SED analysis. We thus leave the analysis of unselected sources for our ongoing work (Lorenzon et al., in prep). All 603 galaxies selected for our SED analysis have available spectroscopic redshifts ($z_{\rm spec}$), distributed over a wide range ($0.1<z<0.6$). 
	
	\section{Panchromatic SED modelling of the data}
	\label{sec:3}

	\subsection{Tools: CIGALE}
	
	We apply full SED (UV+IR) modelling for our QG using the newest release of Code Investigating GALaxy Emission (CIGALE; \citealt{noll09}, \citealt{cigale19}\footnote{\url{https://gitlab.lam.fr/cigale/cigale}}), a state-of-the-art SED modelling and fitting code that combines UV-optical stellar SED with an IR component. For each parameter, CIGALE makes a probability distribution function (PDF) analysis, providing the output value as the likelihood-weighted mean of the PDF (with corresponding error as a likelihood-weighted standard deviation). In the following, we briefly summarise the choice of modules and parameters presented in \hyperref[tab:3.1]{Table 1}. 
	
	\subsubsection*{Stellar component}
	
	We construct the stellar component of our SED model adopting Bruzual $\&$ Charlot stellar population synthesis model (\citealt{bruzual03}, BC03) with a \citealp{chabrier03} IMF. We offer a range of three stellar metallicities closest to that inferred from the mass-metallicity relation for each object.
	
	Because our primary goal is to infer physical characteristics of QGs, we adopt the flexible star-formation history (SFH), which is composed of a delayed component with an additional flexible time at which the star formation is instantaneously affected. Our choice was motivated by recent studies demonstrating that the addition of an extra flexibility in the recent SFH is needed to better recover physical properties (in particular SFR) in a broad range of QGs (e.g.,  \citealt{ciesla16}, \citealt{hunt19},  \citealt{ciesla21}, \citealt{suess22}). The SFH is parametrised as: 
	\begin{equation}
	\label{Eq.1}
	\rm SFR=\begin{cases}
	\mathit{t}\times e^{-t/\tau_{main}}, & \text{when $t\leq t_{\rm trunc}$}\\
	r\times \rm SFR(\mathit{t}), &\text{when $t>t_{\rm trunc}$}
	\end{cases}
	\end{equation}
where $\tau_{main}$ represents the e-folding time of the main stellar population, while $t_{\rm trunc}$ represents the time at which star formation is instantaneously affected (quenched). Parameter $r$ is the ratio between SFRs after quenching and at the moment of quenching, respectively. 
	\begin{table*}
		\caption{Parameters used for modelling the SEDs with CIGALE. All ages and times are given in Gyr.}
		\label{tab:3.1} 
		\centering   
		\scalebox{0.73} {\begin{tabular}{c c c c} 
				\hline
				\toprule 
				\\[-12pt]\\
				Parameter&Values&&Description\\[-7pt]\\\\
				\hline
				& & Star Formation History & \\[-7pt]\\
				
				$\tau_{main}$&1.0, 1.8, 3.0 & & e-folding time (main)\\
				\centering
				Age & 8.5, 9.0, 10.0, 10.5, 11.0, 11.5, 12.0, 12.5, 13.0 & & Population age (main)\\
				Age of quench & [0.01; 3], 10 values log-sampled && Age of the late quenching event\\
				$r$& [0.001, 1], 10 values log-sampled && Ratio of SFR after and prior to quenching\\
				\hline
				& & Stellar emission & \\[-7pt]\\
				
				IMF& Chabrier 2003& & Initial mass function\\
				Z&0.008, 0.02, 0.05& & Stellar metallicity (0.02) in Solar\\ 
				Separation age& 0.01 && Age difference between old and young population\\
				\hline
				& & Dust attenuation & \\[-7pt]\\
				$A^{BC}_{v}$&0.3, 0.8, 1.2, 3.3, 3.8& & V-band attenuation\\    
				slope BC&-0.7& & Power law slope of BC attenuation\\    
				BC to ISM factor&0.3, 0.5, 0.8, 1.0 & & Ratio of the BC-to-ISM attenuation\\             
				slope ISM&-0.7& & ISM attenuation power law slope\\
				\hline
				& & Dust emission & \\[-7pt]\\
				$q_{PAH}$&0.47, 1.12, 3.9, 4.6& & Mass fraction of PAH\\
				$U_{\min}$&1.0, 2.0, 5.0, 10.0, 15.0, 25.0 & & Minimum radiation field\\
				$\alpha$&2.0& & Dust emission power law slope\\
				$\gamma$&0.001, 0.01, 0.1 & & Illuminated fraction\\
				\hline
				& & AGN emission & \\[-7pt]\\
				$r_{ratio}$&60.& &Maximum to minimum radii of the dust torus\\
				$\tau$&1.0, 6.0& &Optical depth at 9.7$\mu$m\\
				$\beta$&-0.5& &Radial dust distribution within the torus\\
				$\gamma$&0.0& &Angular dust distribution within the torus\\
				Opening angle&$100^{\circ}$& &$\gamma$ Opening angle of the torus\\
				$\psi$&$0^{\circ}$, $90^{\circ}$& &Angle between eq.axis and line of sight\\
				$f_{\rm AGN}$&0.0, 0.1, 0.25, 0.5, 0.8&&  AGN fraction\\[-3pt]\\
				
				\hline    
				\bottomrule               
			\end{tabular}
		}\\
		
	\end{table*}
	\subsubsection*{Dust attenuation}
	
	We adopted \cite{charlott00} double power-law attenuation (CF00), which assumes that birth clouds (BCs) and the ISM each attenuate light according to fixed power-law attenuation curves. The formalism is based on age-dependent differential attenuations between young (age $<10^7\:\rm yr$) and old (age $>10^7\:\rm yr$) stars. In this way, the attenuation law intrinsically takes into account variations in the attenuation curves with time. Both attenuations are modelled by a power-law function, providing the amount of attenuation in the $V$ band defined as $\mu=A^{\rm ISM}_{\rm V}/(A^{\rm ISM}_{\rm V}+A^{\rm BC}_{\rm V})$. Following \cite{battisti20} and \cite{ciesla21} we chose to keep both power-law slopes (BC and ISM) of the attenuation fixed at -0.7, and the parameter $\mu$ fixed to 0.3. Because stars older than $<10^7\:\rm yr$ are only attenuated by $A^{\rm ISM}$, we keep parameter $A^{\rm ISM}$ varying between 0.3 and 3, which is demonstrated a good choice for the fitting of quenched objects with detectable dust emission, as recently pointed out by \cite{ciesla21}.
	
	\subsubsection*{Dust emission}
	
	For modelling the galaxies' IR SEDs we adopt the physically motivated dust library of \citealp{dl07} (DL07 hereafter). In DL07, IR SEDs are calculated for dust grains heated by starlight for various distributions of intensities. The majority of the dust is heated by a radiation field with constant intensity from the diffuse ISM, while muchj smaller fraction of dust ($\gamma$) is exposed to starlight with interstellar radiation field (ISRF) intensity in a range between $U_{\rm min}$ to $U_{\rm max}$ following a power-law distribution. We fixed the maximum radiation field intensity to $U_{\rm max}=10^6$ and sample different $U_{\rm min}$, keeping the emission slope fixed at $\beta=2$. The fraction of the $M_{\rm dust}$ in the form of PAH grains ($q_{\mathrm{PAH}}$) was sampled within the range ($0.47< q_{\mathrm{PAH}}<4.6$). We log-sample illumination fractions ($\gamma$) between $0.001$ and $0.1$ with the step of 0.01. As showed in \cite{berta16}, sampling towards larger $\gamma$ may increase the chance of overestimating $M_{\rm dust}$. In our modelling, $L_{\rm IR}$ is an integral of a SED over the rest-frame wavelength range of $\lambda=8-1000\:\mu$m, while $M_{\rm dust}$ is derived by fitting and normalising the IR photometry to the DL07 library. 
	
	\begin{figure*}[ht]
		\vspace{-0.2cm}
		\centering
		\includegraphics [width=17.33cm]{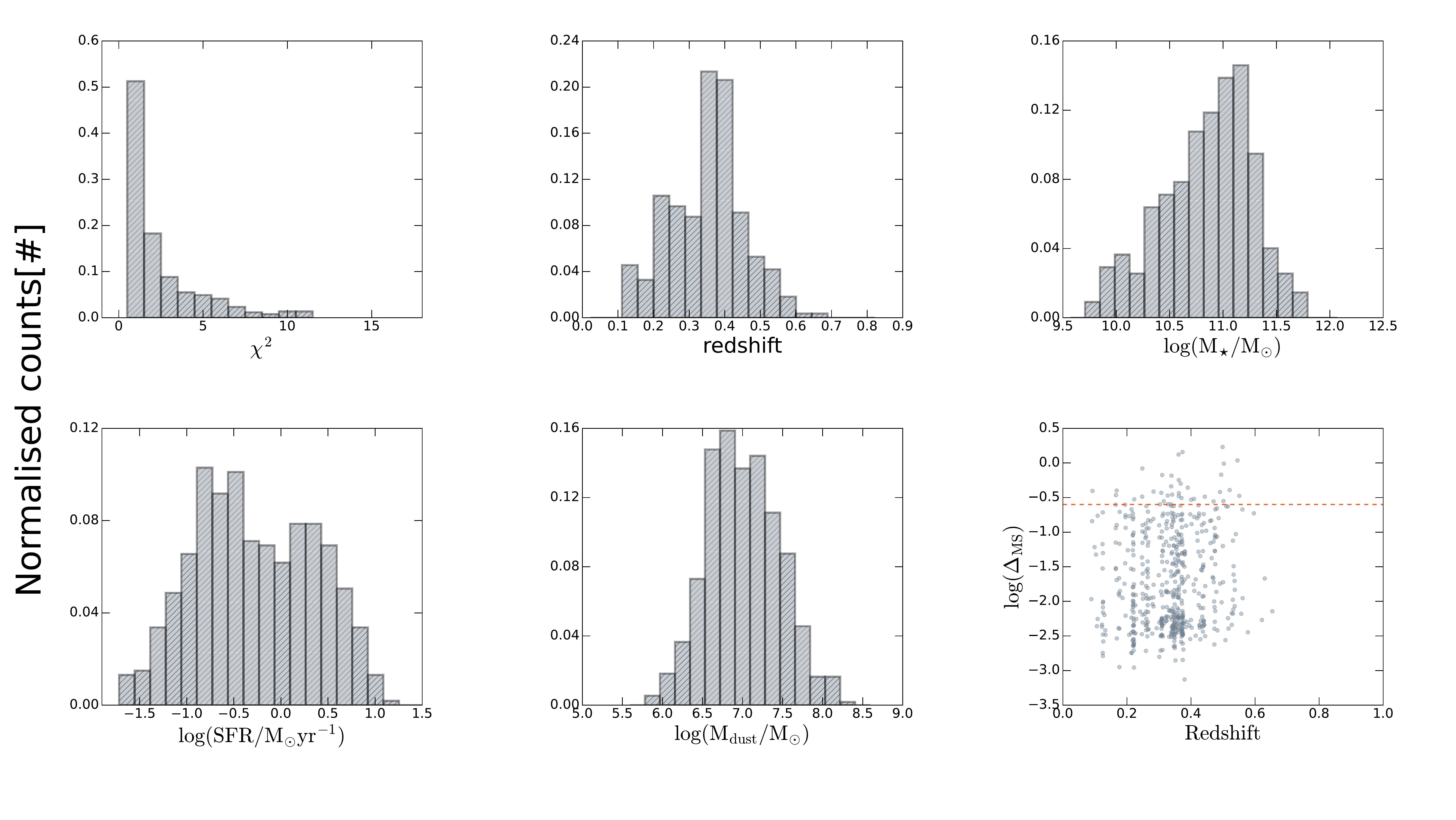}
		\caption{Distributions of the physical properties estimated for our candidate quiescent galaxies from the SED fitting with CIGALE. \texttt{Clockwise, from top left to bottom right}: Goodness of fit expressed as reduced $\chi^{2}$; galaxy redshift; stellar mass; star-formation rate (SFR), dust mass and linear offset of the galaxy’s observed SFR to the SFR expected from the modelled MS ($\Delta_{\rm MS}$ in $\log$ scale) as a function of redshift.}
		\label{fig:Fig.1}
	\end{figure*}
	
	We also chose to derive the fractional contribution of the AGN, defined as the relative impact of the dusty torus of the AGN to the $L_{\rm IR}$ ("AGN fraction"). We adopt AGN templates presented in \citealp{fritz06} (see also \citealt{feltre12}). 
	The parameters in the AGN model were matched to those from \cite{donevski20}. Due to computational reasons we somewhat reduce the number of input options, and model the two extreme values for inclination angle ($0^{\circ}$ and $90^{\circ}$).
	

	\subsubsection*{Constraints on the parameters}
	
	We fit the full datasets with the models defined in previous section. To do so, we perform the full SED (UV-to-FIR) modelling, but chose the option in CIGALE to predict $D_{\rm n}4000$. We did several runs of fits until we obtain a good match between the modelled and the observed $D_{\rm n}4000$. This procedure is important and ensures that our choice of modelling parameters with the inclusion of dust correctly matches the galaxy ages and SFH, which is important to overcome possible degeneracies. 
	Before using our SED-derived quantities for the science analysis, we confirm that all fitted SEDs are of a good quality, indicated with a small average $\chi^{2}$  (median is find to be $\chi^{2}=1.12\pm0.33$). We also assign the modelling option available within CIGALE to produce mock catalogues, then following the approach implemented by \cite{malek19} and \cite{donevski20} to verify that our SED fitting procedure does not introduce significant systematics to our dust-related measurements, in particular $M_{\rm dust}$ (see \hyperref[sec:appB]{Appendix B}). 
	
	\subsection{Statistical properties of our sample}
	From the further analysis we exclude AGN powered objects based on their SED output (55/603 sources with $f_{\rm AGN}>0.25$, or $9\%$ of the total sample).\footnote{We also double-check for additional X-ray-bright AGNs in the COSMOS (\citealt{civano16}) and find that  $\sim 30\%$ of AGNs we identify via SED modelling also have excess X/radio emission.} After this step, the remaining 548 sources are used for our final analysis. We stress here that none of QGs from our final sample is classified as "star-forming" based on BPT diagnostics, while very  minor (18/548) number of objects are classified as "composite".\footnote{Recall that these classifications are based on 1.5" diameter fiber spectra. Thus, the "non-classifiable" sample likely includes galaxies which do not have star formation in the central bulge but may have residual star formation in the disk.} 
	
	In \hyperref[fig:Fig.1]{Fig. 1,} we show the distribution of SED derived properties. We infer the median redshift of $z=0.37$ with the corresponding 16-84th percentile range ($z=0.26-0.48$). As expected for evolved objects, QGs from our hCOSMOS sample are very massive systems ($\log(\rm M_{\star}/M_{\odot})=10.92^{+0.24}_{-0.32}$), The median spectral age indicator is $D_{\rm n}4000=1.73$, while the median and interquartile range of IR luminosity is $\log(L_{\rm IR}/L_{\odot})=10.28^{+0.30}_{-0.42}$, which is two orders of magnitude lower than in DSFGs at a given redshift and stellar mass range. We apply the functional form of the MS defined by \citealp{speagle14} (their 'best-fit', provided with Eq. 28) to check the position of our QGs with respect to the MS. With this check we unveil that $88\%$ of dusty QGs from our sample reside well bellow the MS, with the linear offset to the SFR expected for MS objects being $\log(\Delta_{\rm MS})\leq-0.6$. Remaining $\sim 12\%$ of sources can be considered galaxies approaching "green valley" or MS. 
	
We also check if the results presented in the rest of the text hold if  alternative selection criteria for QGs are applied. In \hyperref[App:A3]{Appendix C} we confront different tracers for galaxy quiescence (both on the basis of rest-frame colours and spectroscopy) and demonstrate that selection have little or no impact on findings presented in the next sections.
	
		\begin{figure*}[h]
		\centering
		\includegraphics [width=10.69cm]{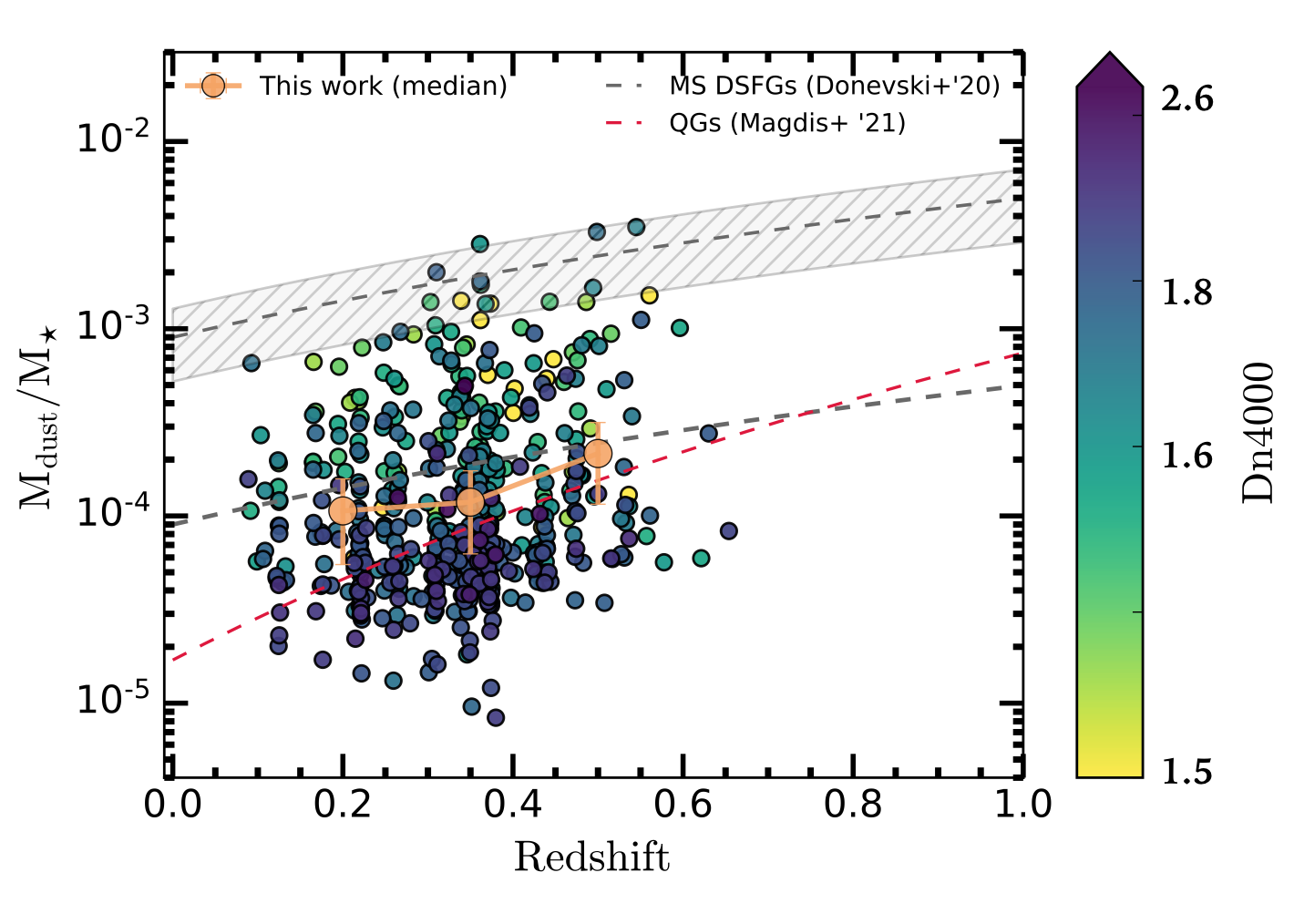}
		
		\caption{Observed redshift evolution of $M_{\rm dust}/M_{\star}$ in dusty QGs from this work. Individual values are displayed with circles, coloured with corresponding $D_{\rm n}4000$. Binned medians and associated uncertainties  (16th-84th percentile range) are shown with sandy brown circles and lines, respectively. For comparison, 
			we show the best fit from the stacking analysis of \cite{magdis21} (red dashed line). The dark-grey line and shaded area describe the modelled evolution of ALMA detected MS galaxies with a functional form of $M_{\rm dust}/M_{\star}\propto  k\times (1+z)^{2.5}$ (\citealt{donevski20}), while the dark-grey dashed line shows the shift of 1 dex of this MS scaling relation.}
		\label{fig:Fig.2}
	\end{figure*}	

	\section{ISM diagnostics of the general population of QGs from hCOSMOS}
	\label{sec:4}
	
	\subsection{Evolution of $M_{\rm dust}/M_{\star}$ with redshift}
	
	In order to gain insight into evolution of galaxy dust content in our QG, we apply $M_{\rm dust}/M_{\star}$ as a tool to assess the efficiency of the specific dust production and destruction mechanisms. From our multi-band SED fitting, we find the median of $M_{\rm dust}/M_{\star}=1.5^{+0.6}_{-0.7}\times10^{-4}$. This is more than an order of magnitude lower than what has been estimated for MS DSFGs within the same redshift and stellar mass range (e.g., \citealt{daCunha10}, \citealt{driver18}), and is two orders of magnitude lower than in dusty starbursts (\citealt{dacunha15}, \citealt{donevski20}, \citealt{pantoni21}). As seen in \hyperref[fig:Fig.2]{Fig.2,}, where we show the redshift evolution of the $M_{\rm dust}/M_{\star}$ for the full sample, some of our dusty QGs are approaching the main-sequence (MS) relation modelled for ALMA identified dusty galaxies at $z<1$ (see the Section 4.4 in \citealt{donevski20} for details). 
	That said, a handful of QGs in our sample have $M_{\rm dust}/M_{\star}$ comparable to those of MS DSFGs albeit having lower $L_{\rm IR}$ on average. The median values of our QGs agrees reasonably well with the modelled cosmic evolution of $M_{\rm dust}/M_{\star}$ in QGs at $0<z<1$, obtained from stacking IR-to-radio maps of colour-selected, massive ($M_{\star}>10^{10}M_{\odot}$) galaxies in the COSMOS field (\citealt{magdis21}). 
The average $M_{\rm dust}/M_{\star}$ from \cite{magdis21} are 0.2-0.5 dex bellow the medians from this work, which is expected due to the stacking technique they used to reach fainter $L_{\rm dust}$. Nevertheless, our data agree within $1\sigma$, yet suggesting slightly milder decline with $z$. As radio-mode AGN feedback is expected to dominate the faster decline of $M_{\rm gas}$ (and supposedly $M_{\rm dust}$; \citealt{feldman15}, \citealt{wu18}), the exclusion of candidate AGNs from our final sample likely contribute to the shallower evolutionary trend with respect to those inferred by \cite{magdis21}. In the accompanying paper (Lorenzon et al. in prep.), we stacked the FIR maps of all unselected QGs from the hCOSMOS parent sample, and obtain the upper limits of $M_{\rm dust}/M_{\star}=1.4\times10^{-5}$ and $M_{\rm dust}/M_{\star}=3\times10^{-5}$, for $z=0.25$ and $z=0.37$ respectively. These values, as well as their redshift evolution, are very similar to that of \cite{magdis21}, implying a less dramatic cosmic evolution of the ISM in dusty QGs in comparison to the general population of QGs.  
	
	In \hyperref[fig:Fig.2]{Fig. 2} we see that galaxies with higher $D_{\rm n}4000$ generally attain lower specific dust mass than those with smaller $D_{\rm n}4000$. Because $D_{\rm n}4000$ is a proxy for stellar population age, more evolved systems with older stellar populations are naturally expected to have lower $M_{\rm dust}/M_{\star}$ (e.g., \citealt{hjorth14}, \citealt{pantoni19}, \citealp{michalowski19}). 
	However, \hyperref[fig:Fig.2]{Fig. 2} reveals that even for objects observed at the same redshift, there is a significant spread in $M_{\rm dust}/M_{\star}$ which does not always coincide with $D_{\rm n}4000$. This could suggest diverse ISM conditions in our QGs, (i.e., large variability of molecular gas fractions, recent merger activities etc.) which may result in different dust production/destruction timescales. This slightly challenges the interpretation by \cite{magdis21}, who argue that prompt decline of $M_{\rm dust}/M_{\star}$ must be due to quick gas removal with no further replenishment (i.e., due to minor mergers or accretion from the cosmic web).

	\subsection{Evolution of $M_{\rm dust}/M_{\star}$ with sSFR}
	
	\begin{figure*}[h]
		\centering
		\includegraphics [width=15.89cm]{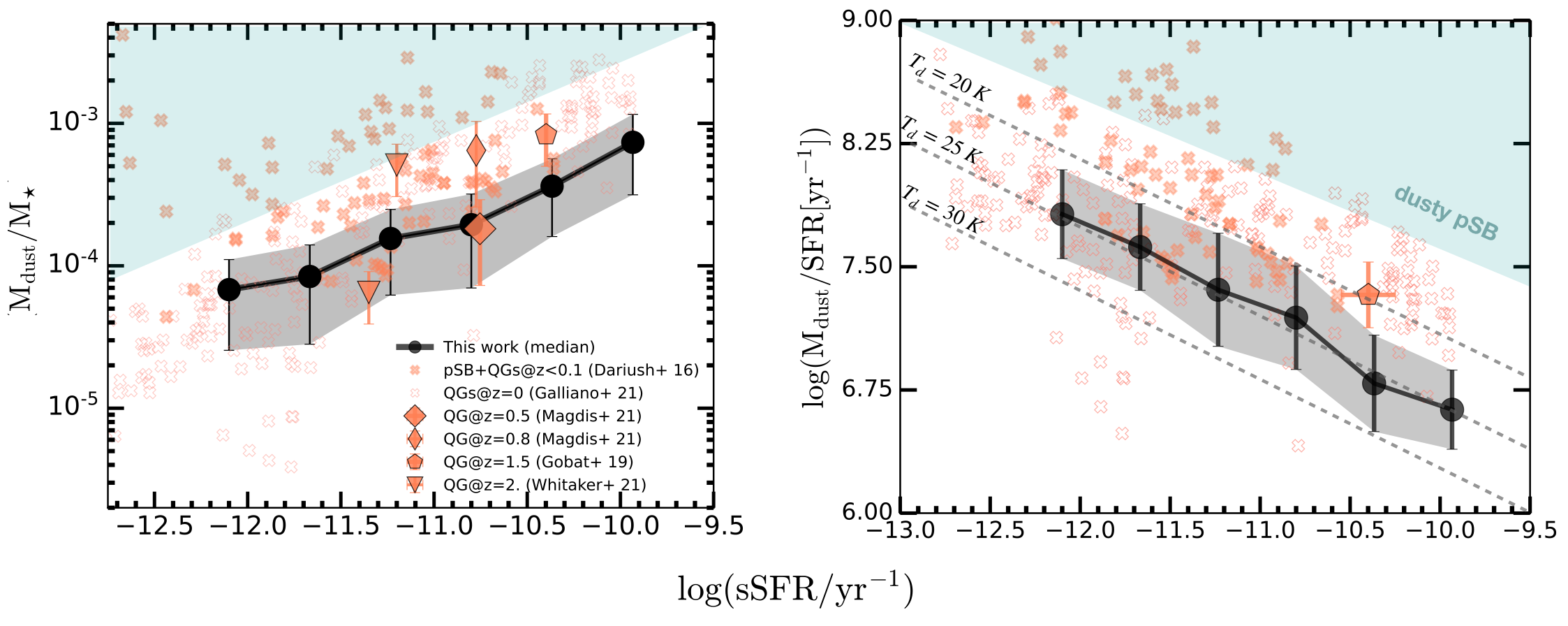}
		\caption{Evolution with sSFR for $M_{\rm dust}/M_{\star}$ (\textit{left}) and  $M_{\rm dust}/\rm SFR$ (\textit{right}). \texttt{Left panel:}The binned medians and corresponding 16th–84th percentiles of this sample are presented with black circles and with dark grey shaded area, respectively. Open crosses indicate the subsample of quiescent objects drawn from the Dustpedia archive (\citealt{galliano21}). Filled crosses show the compilation of dusty QGs and pSB galaxies at $z<0.1$ detected in H-ATLAS (\citealt{dariush16}). Displayed with salmon-coloured symbols are estimates for the stacked and individual QGs at higher-redshifts: stacks are represented with diamonds and pentagons (\citealt{gobat18}, \citealt{magdis21}), while the individual detections are annotated with triangles (\citealt{whitaker21a},). In both panels the cyan-shaded region illustrates the parameter space where majority of dusty pSB galaxies detected with ALMA reside (\citealt{lipsb19}); \texttt{Right panel:} The symbols are of the same meaning as in the left panel. Dashed lines roughly track the position of sources within different $T_{\rm dust}$, calculated by scaling the strength of the ISRF intensity as derived from CIGALE.
		}
		\label{fig:Fig. 3}
	\end{figure*}
	
	In star-forming galaxies, dust mass is expected to be a good tracer of the $M_{\rm gas}$ (\citealt{daCunha10}, \citealt{santini14}, \citealt{rowlands15}, \citealt{scoville17b}, \citealt{kirkpatrick17}, \citealt{aoyama19}), while $M_{\rm gas}$ and SFR are linked through the known Kennicutt–Schmidt relation (KS, \citealt{schmidt59}, \citealt{kennicutt98}, \citealt{sargent14}). It is thus important to unveil how $M_{\rm dust}$, $M_{\star}$ and SFR are related in dusty QGs at intermediate redshifts.
	
	In \hyperref[fig:Fig. 3]{Fig. 3,} we show two ISM diagnostics as a function of sSFR: $M_{\rm dust}/M_{\star}$ (left panel), and $M_{\rm dust}/\rm SFR$ (right panel). Both relations extend over two orders of magnitude, with standard deviations of $\sim 0.4-0.5\: \rm dex$ in each sSFR bin. This hints to complexity of ISM conditions in dusty QGs, similar to those seen in high-$z$ DSFGs (e.g., \citealt{dacunha15}, \citealt{calura17}, \citealt{strandet17}, \citealt{as2uds19}, \citealt{donevski20}). Such significant spread in $M_{\rm dust}/M_{\star}$ and $M_{\rm dust}/\rm SFR$ is interesting to understand, as QGs are expected to share similar, redshift independent SEDs, consequence of almost constant dust temperature ($T_{\rm dust}\sim20$ K) and less turbulent ISM as compared to DSFGs (\citealt{andreani18}, \citealt{magdis21}). If we adopt our SED derived $\langle U\rangle_{\rm min}$ and convert it to $T_{\rm dust}$ via calibration given by \cite{schreiber18}, we obtain values ranging from 17K to 33K, with the median and absolute median deviation of $24\pm2.3$ K. This $T_{\rm dust}$ is $\sim4$ K warmer than the average value found by \cite{magdis21}. Variations in $T_{\rm dust}$ are seen in each sSFR bin, in agreement with \cite{martis19} and \cite{nersesian19} who reveal the similarly wide range of values among the dusty galaxies with low sSFR. Differences in $T_{\rm dust}$ can be due to the dust grains being exposed to different ISRFs (\citealt{nersesian19}), but can also be related to intrinsic properties of dust, such that galaxies with the higher fraction of larger (smaller) grains produce cooler (warmer) SEDs (\citealt{relano22}, \citealt{nishida22}). This, again, refers to diverse channels for the ISM evolution in observed QGs.
	
	The relations shown in \hyperref[fig:Fig. 3]{Fig. 3} can provide insight into the evolutionary stages of our dusty QGs. For example, the trend of $M_{\rm dust}/M_{\star}$ with $\rm sSFR$ can be partially interpreted as age-evolutionary sequence (\citealt{calura17}), such that QGs from the upper-right side of the diagram could be dominated by objects that have younger stellar age and higher gas fraction ($f_{\rm gas}=M_{\rm gas}/(M_{\star}+M_{\rm gas})$). The subsequent decrease of sSFR is than mostly due to exhaustion of their gas reservoirs, reflecting the efficiency of ISM removal (\citealt{gobat20}). The anti-correlation of $M_{\rm dust}/\rm SFR$ and sSFR is often viewed either as a proxy for the metallicity-dependent gas depletion timescale ($M_{\rm dust}/\rm SFR \propto \tau_{\rm dep} (Z/Z_{\odot}$), e.g., \citealt{bethermin15}, \citealt{magdis21}), or inverse of the mean radiation field ($\left\langle U_{\rm mean} \right\rangle$, e.g., \citealt{hunt14}, \citealt{martis19}). The former implies that galaxies with smaller $M_{\rm dust}/\rm SFR$ and high sSFR can be compatible with shorter gas depletion timescales (i.e., $\tau_{\rm dep}\sim 500$ Myr -1 Gyr), indicative of MS DSFGs (\citealt{bethermin15}, \citealt{scoville17}, \citealt{liu19b}). Consequently, objects that populate upper left part of $M_{\rm dust}/\rm SFR$-sSFR plane are expected to have longer depletion, order of few Gyr. 
	
Interestingly, the literature lacks systematically selected statistical datasets of dusty QGs that would be fully suitable for comparison with our sample. Therefore, we chose to compare our data to the widely used benchmark samples of dusty galaxies in the nearby universe, such as the H-ATLAS sample of optically red, dusty galaxies (\citealt [hereafter \texttt{D16}]{dariush16}), and the Dustpedia archive (\citealt [hereafter \texttt{G21}]{galliano21}; see also \citealt{lianou19}, \citealt{nersesian19}). The D16 sample comprises of 78 galaxies at $z<0.1$ that are selected based on $>5\sigma$ \textit{Herschel} detection at $250\:\mu$m, and for which SDSS counterparts have red colours based on $NUV-r$ diagnostics. The sample from G21 consists of $\sim 770$ very local galaxies with angular sizes of $D_{25}> 1^{\prime}$ and similarly significant \textit{Herschel} detections. The majority of Dustpedia sources are SF galaxies, and for comparison we only chose those sources (180 in total) within the same range of sSFR and $M_{\star}$ as our QGs ($\log(\rm sSFR/\rm yr^{-1})<-10$ and $M_{\star}>10^{10}M_{\odot}$). However, this only allows very rough comparison because Dustpedia galaxies are not specifically selected as quiescent, i.e. based on their $D_{\rm n}4000$. On top of this, the median stellar mass of G21 sample is $\log(M_{\star}/M_{\odot})=10.4$, which is $\sim0.5$ dex below the median of our QGs. Therefore, we stress that any conclusion resulting from comparisons with these two samples should be taken with caution. 
Nevertheless, as can be derived from both panels of \hyperref[fig:Fig. 3]{Fig. 3}, our estimates follow the similar evolutionary shape with sSFR as galaxies from G21 and D16, while the derived dust related properties differ among the samples. The points from G21 depart from our median $M_{\rm dust}/M_{\star}$ and $M_{\rm dust}/\rm SFR$ at higher sSFR, while those from D16 lie systematically above our QGs over the entire sSFR range. 
Although all three works applied different fitting methods and model assumptions\footnote{In G21 the CIGALE was used for obtaining $M_{\star}$ while $M_{\rm dust}$ is inferred by the code HerBIE (\citealt{galliano18}) which adopts the framework of THEMIS dust model (\citealt{themis17}). The properties of D16 are derived by MAGPHYS (\citealt{dacunha08})}, it is more likely that diversity in values is due to physics and selection bias. The $T_{\rm dust}$ derived in G21 has a median of $\sim19\rm K$, which is $\sim5$K colder than average $T_{\rm dust}$ of our sample. Therefore, the difference with G21 seen towards higher sSFR likely arises due to G21 galaxies being less massive and more colder than our QGs. The average $M_{\star}$ of D16 galaxies, on the other hand, is similar to ours ($\log(M_{\star}/M_{\odot})=10.8$), but their $M_{\rm dust}$ are systematically higher. This can be a consequence of selection bias and noise effects. The H-ATLAS survey covers an area that is $\sim100$ times larger but $\sim5$ times shallower than the COSMOS field. Hence, $>5\sigma$ selection would return the most extreme objects that are much brighter and "dustier" than our QGs. In addition, noise effects can impact $M_{\rm dust}$ estimation (i.e., \citealt{berta16}). As a consequence, most objects from D16 enter the region of ALMA-observed post-SB galaxies (\citealt{lipsb19}, marked as the cyan shaded area in \hyperref[fig:Fig. 3]{Fig. 3}). 

The median trend of our specific dust masses agree within $1\sigma$ with those derived for QGs at higher-$z$ ($0.5<z<2$; \citealt{gobat18}, \citealt{magdis21}, \citealt{whitaker21a}). These studies reach somewhat contrasting conclusions regarding the physical properties of detected objects. \cite{magdis21} argue of long $\tau_{\rm dep}$ (>Gyr) and roughly constant $T_{\rm dust}\sim20$ K, whereas the two QGs from \cite{whitaker21a} are characterised by shorter depletion timescales ($\tau_{\rm dep}\sim 100$ Myr), and very different $T_{\rm dust}$ (14 K and 34 K, respectively), and $f_{\rm gas}$ ($4.6\pm0.5\%$ and $0.6\pm0.1\%$, respectively). The dusty QG identified by \cite{gobat18} is found to host a larger cold gas fraction ($5-10\%$) under dust temperature of $T_{\rm dust}=20$ K. \cite{whitaker21a} suggested that a large range in specific dust masses hints to diverse evolutionary routes to quiescence. Therefore, it is of vital importance to investigate what causes the large dispersion in $M_{\rm dust}/M_{\star}$ seen in QGs. We thoroughly explore this question in the next Sections. 

	\begin{figure*}[ht]
		\label{Fig:app3}
		\centering
		\includegraphics [width=14.33cm]{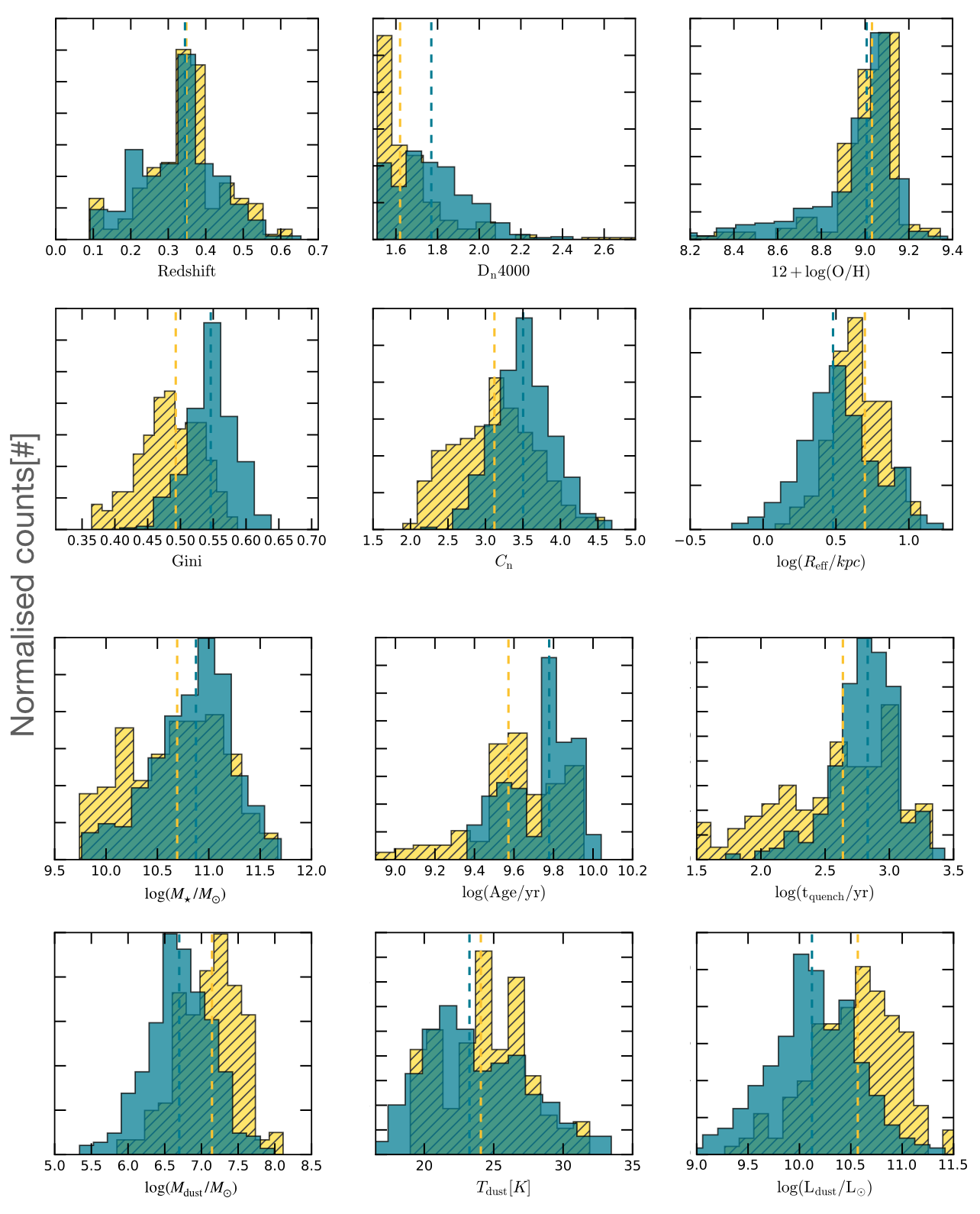}
		\caption{Structural and SED-derived physical properties of QGs from hCOSMOS. Normalised distributions of values for various physical quantities are displayed for two morphological categories; quiescent spirals and ellipticals (sQGs and eQGs; yellow and dark cyan colours, respectively). The first row delineates properties obtained from the Hectospec spectroscopy (redshift, $D_{\rm n}4000$ and gas-metallicity (as oxygen abundance) expressed as $\rm 12+\log(\rm O/H)$). The second row displays some of the main structural properties derived from HST images observed via F814W filter (Gini index, concentration and effective radius in kpc) from the catalogue of \citealt{cassata09}). The third and fourth row show various physical properties estimated from multiwavelength SED modelling of our QGs with the code CIGALE. These are, from upper left to lower right, respectively: $\log(M_{\star}/M_{\odot})$, $\log(\rm Age/yr)$, $\log(\rm t_{\rm quench}/yr)$, $\log(M_{\rm dust}/M_{\odot})$, $T_{\rm dust}$  and  $\log(L_{\rm dust}/L_{\odot})$.}
	\end{figure*}

	\begin{figure*}[h]
		\centering
		\includegraphics [width=17.69cm]{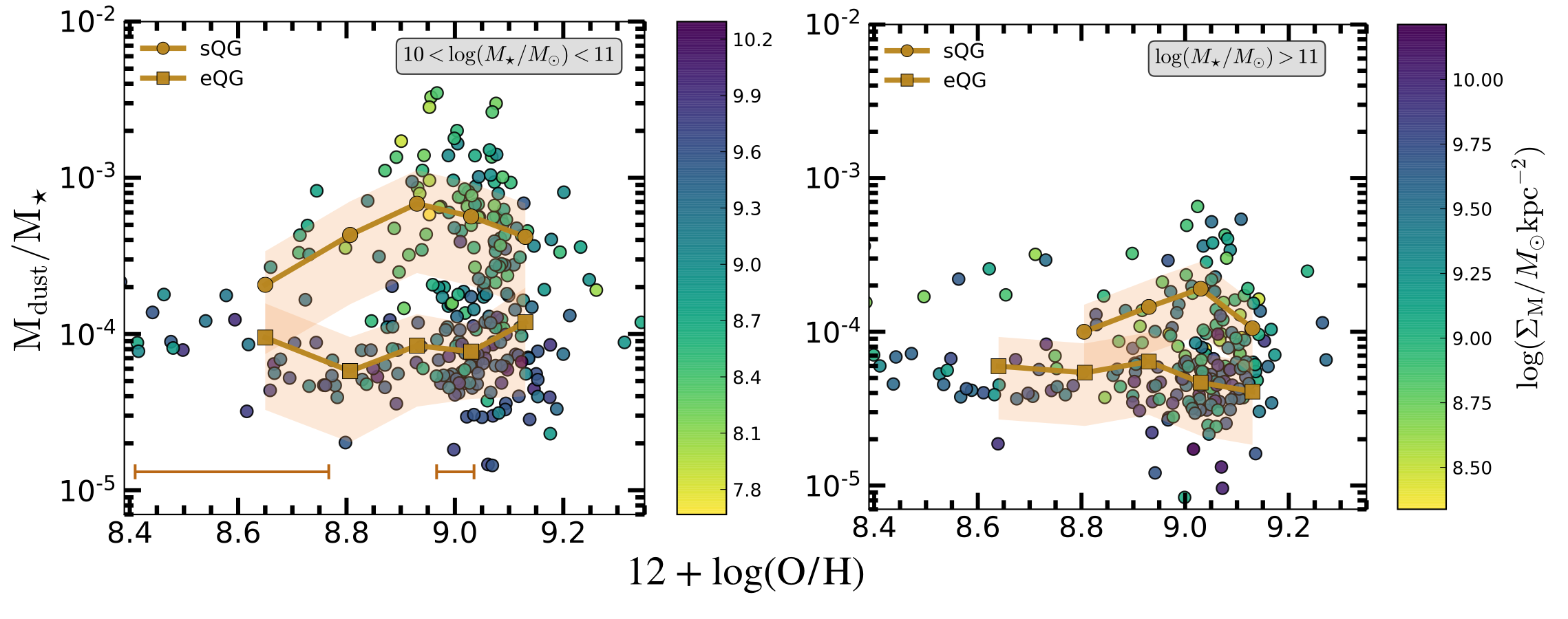}

		\caption{\texttt{Upper panel:} Evolution of $M_{\rm dust}/M_{\star}$ as a function of gas-phase metallicity, colour-coded for the stellar mass surface densities (expressed on a logarithmic scale). Galaxies are divided in two mass bins: $\log(M_{\star}/M_{\odot})<11$ (\textit{left}), and $\log(M_{\star}/M_{\odot})>11$ (\textit{right}). Brown markers and shaded regions represent the binned medians and 16th-84th percentile range of the resulting distribution for morphologically classified sub-samples of dusty QGs: those classified as elliptical-type QGs (eQGs; squares) and spiral QGs (sQGs; circles). The typical error in gas-metallicity estimates for solar and super-solar values are displayed with horizontal lines, respectively.
		}
		
		\label{fig:Fig.4}
	\end{figure*}
	
	
	\section{Morphological impact on the evolution of dust abundance in QGs}
	\label{sec:5}
	
	Galaxy morphology is often directly linked to some of the dust-related properties of galaxies (e.g., \citealt{smith14}, \citealt{nersesian19}, \citealt{casasola20}). Our goal in this Section is to explore if the observed evolution of specific dust mass at intermediate redshifts is driven by the morphological type of our galaxies. For this reason we adopt the morphological classification by \cite{tasca09}. They infer morphological types of galaxies in COSMOS by applying the automated method on parameters such as Concentration ($C_{\rm n}$), Gini index ($G$), Asymmetry and M20 moment, calculated from the HST F814W imaging (\citealt{cassata09}). With this we split the full sample of our QGs into two sub-classes: elliptical QGs (eQGs; 385 objects out of 548; $70\%)$ and spiral QGs (sQGs; 163 objects out of 548; $30\%$)\footnote{We note that morphological classification remains almost unchanged when using other methods, i.e., Zurich Estimator of Structural Types (ZEST) by \cite{scarlata07}}. The distributions of their main structural and physical properties are presented in \hyperref[Fig:app3]{Fig. 4}. Through the rest of the paper, we extensively investigate evolution of specific dust mass with different parameters in these two morphological groups.

	\subsection{The relationship between $M_{\rm dust}/M_{\star}$ and gas-phase metallicity}
	\label{sec:5.1}
	
	Gas-phase metallicity is considered one of the crucial parameters influencing the dust life cycle in galaxies (e.g., \citealt{asano13}, \citealt{pantoni19}, \citealt{hou19}, \citealt{triani20}). The unique aspect of our hCOSMOS dataset is a combination of self-consistently SED-derived $M_{\rm dust}$ and $M_{\star}$, and independent measurements of $Z_{\rm gas}$. This enables us to investigate, for the first time, how the $M_{\rm dust}/M_{\star}$ evolves as a function of $Z_{\rm gas}$ in two different morphological categories of QGs beyond the local Universe. 
	The majority of our objects (76\%) have "super-solar" $Z_{\rm gas}$ with the median that reaches twice the solar value ($12+\log(\rm O/H)=9.01^{+0.07}_{-0.13}$). Both distributions and medians of $Z_{\rm gas}$ are similar between the two morphological groups (see \hyperref[Fig:app3]{Fig. 5}). In \hyperref[fig:Fig.4]{Fig. 5}. we showcase the relation of $M_{\rm dust}/M_{\star}$ with $Z_{\rm gas}$ for two stellar mass bins, colour-coding their stellar mass surface densities (defined as $\Sigma_{\rm M}=M_{\star}/2\pi R_{\rm eff}^{2}$, where $R_{\rm eff}$ is effective radius in kpc based on the measurements from \citealt{cassata09}). 
	
	The \hyperref[fig:Fig.4]{Fig. 5} reveals several interesting features: \textbf{(1)} At fixed $Z_{\rm gas}$ the observed $M_{\rm dust}/M_{\star}$ varies amongst morphological types. In sQGs $M_{\rm dust}/M_{\star}$ is systematically higher than in eQGs, and it evolves in a more complex way. When transiting from (sub-)solar to the super-solar gas metallicities, sQGs undergo turnover of the relation between $M_{\rm dust}/M_{\star}$ and $Z_{\rm gas}$. Such behaviour is visible in both mass bins, being more prominent for $10<\log(M_{\star}/M_{\odot})<11$. We note that the trend observed in sQGs should be consider rather tentative, as typical uncertainties in $Z_{\rm gas}$ measurements are quite large ($\sim 0.4$) for $12+\log(\rm O/H)\leq 8.8$. Irrespectively of this, eQGs maintain remarkably flat $M_{\rm dust}/M_{\star}$  over range of metallicities. To our knowledge, this is the first time that the morphological dependence in $M_{\rm dust}/M_{\star}$ with $Z_{\rm gas}$ has been observed in QGs. This result strongly indicates that morphology-type impact on the dust content extends at least to $z\sim0.6$, complementing the conclusions from known studies of QGs at $z\sim0$ (\cite{smith14}, \citealt{beeston18}, \citealt{nersesian19}, \citealt{casasola22}); \textbf{(2)} For $12+\log(\rm O/H)\sim9$ there is a range of specific dust masses that extends over $\sim 2$ orders of magnitude and exceeds the median deviation in both morphological types. This implies that our QGs are subject to complex interplay of processes contributing to the dust life cycle. 
	
	In the first place, dust particles can be internally produced via core-collapse supernovae (CC SNe) and in outflows of AGB stars, and can further grow via collisions and accretion of free metals in the ISM. The absence of a clear trend in \hyperref[fig:Fig.4]{Fig. 5} suggests that balance between formation and destruction of dust grains is altered at super-solar $Z_{\rm gas}$. Specific dust masses in eQGs saturates around $M_{\rm dust}/M_{\star}\sim 10^{-4}$ over range of metallicities. Interpreting such behavior is not trivial. For evolved galaxies, the $M_{\rm dust}/M_{\star}$ is expected to arise from the "competition" between the dust-to-gas ratio ($\delta_{\rm DGR}$) increasing and $f_{\rm gas}$ decreasing (i.e., \citealt{asano13}, \citealt{bethermin15}). By using resolved observations of M101 galaxy, \cite{vilchez19} find constant $M_{\rm dust}/M_{\star}$ with $Z_{\rm gas}$ in the outer part characterised by sub-solar $Z_{\rm gas}$. They interpret such flat trend as a consequence of constant yield ratio dominated by stellar sources under constant gas fraction ($f_{\rm gas}\sim0.1$). Nevertheless, such scenario seems unlikely for our sample, as it is favoured for sub-solar environments, but not for the metal-rich ISM. Furthermore, cold gas fraction in our QGs is expected to vary, consequence of the cosmic evolution. If the dust yield is solely by stellar sources and subsequently removed by efficient outflows, one would expect strong anti-correlation between $M_{\rm dust}/M_{\star}$ and $Z_{\rm gas}$, as known in most of local, DSFGs (i.e, \citealt{casasola22}). This contradicts our data and suggests that prolonged dust productions of varying efficiencies may be required to balance destructive processes responsible for gas (and subsequently, dust) decline. 
	
In this regard, growing number of theoretical studies propose the grain growth of dust in ISM to play important role in metal-rich galaxies (e.g., \citealt{asano13}, \citealt{remyruyer14}, \citealt{hirashita15}, \citealt{zhukovska16}, \citealt{deVis19}, \citealt{aoyama19}). In such scenario, if $Z_{\rm gas}$ in a galaxy exceeds a certain critical value which strongly depends on galaxy star-formation history, the grain growth becomes active and the $M_{\rm dust}$ rapidly increases until metals are depleted from the ISM. The evolution of specific dust mass in our sQGs qualitatively agrees with the predictions from the grain growth scenario. $M_{\rm dust}/M_{\star}$ first rises up to certain $Z_{\rm gas}$, and levels off when most of the available metals are depleted onto the dust grains. At the same time, global dust destruction should decrease available cold gas, reducing the material for the further dust production. This interpretation was proposed in \cite{casasola20} who explore morphological impact on the scatter in dust scaling relations with $Z_{\rm gas}$ in Dustpedia galaxies. They invoke effect of grain growth to explain observed specific dust masses ($M_{\rm dust}/M_{\star}>0.0001$) and high average dust-to-gas ratio ($\delta_{\rm DGR}\gtrsim 1/100)$) in objects with super-solar $Z_{\rm gas}$. In principle, expected accretion time for grain growth scales to the inverse of the cold gas fraction, being very rapid ($\sim 10^7-10^8$ yr) in QGs with $f_{\rm gas}>0.1$ (\citealt{hirashita15}, \citealp{hirashita17}). With this production channel being dominant, the observed scatter in $M_{\rm dust}/M_{\star}$ at a given $Z_{\rm gas}$ may reflect the variation of growth efficiencies. The significance of such process can be traced via dust depletion to metals (quantified as "dust-to-metal ratio", $\delta_{\mathrm{DTM}}$, \citealt{deVis19}), which we explicitly investigate with models in \hyperref[sec:6.4]{Section 6.4}.
	
Our QGs span a wide range of optical sizes, with stellar mass surface densities (defined as $\Sigma_{\rm M}=M_{\star}/2\pi R_{\rm eff}^{2}$) extending over $\gtrsim2$ orders of magnitude. From \hyperref[fig:Fig.4]{Fig. 5} we see that over range of metallicities, dusty QGs with lower surface densities tend to have larger $M_{\rm dust}/M_{\star}$. This may suggests that processes driving the size growth in the QGs may leave observational imprints on their relative dust abundance. Because the presence of a companion galaxy can affect the dust scaling relations, several works argue that i.e., extended sizes of dusty QGs can support the idea that some fraction of dust and metals is acquired via external sources, most likely via mergers with dust-rich satellites (\citealt{rowlands12}, \citealt{martini13}, \citealt{lianou16}, \citealt{dariush16}, \citealt{kokusho19}). We will return to the connection with surface densities in the sections where we examine implications related to the channels of dust production and dust removal, and their link with post-quenching timescales (\hyperref[sec:5.4]{Section 5.4} and \hyperref[sec:5.5]{Section 5.5}). The comprehensive theoretical interpretation based on the models will be presented in \hyperref[sec:6]{Section 6}. 
	
	\subsection{Dissecting the age-dependent evolution of $M_{\rm dust}/M_{\star}$}
	\label{sec:5.2}
	
\begin{figure}[h]
		\centering
		\includegraphics [width=9.19cm]{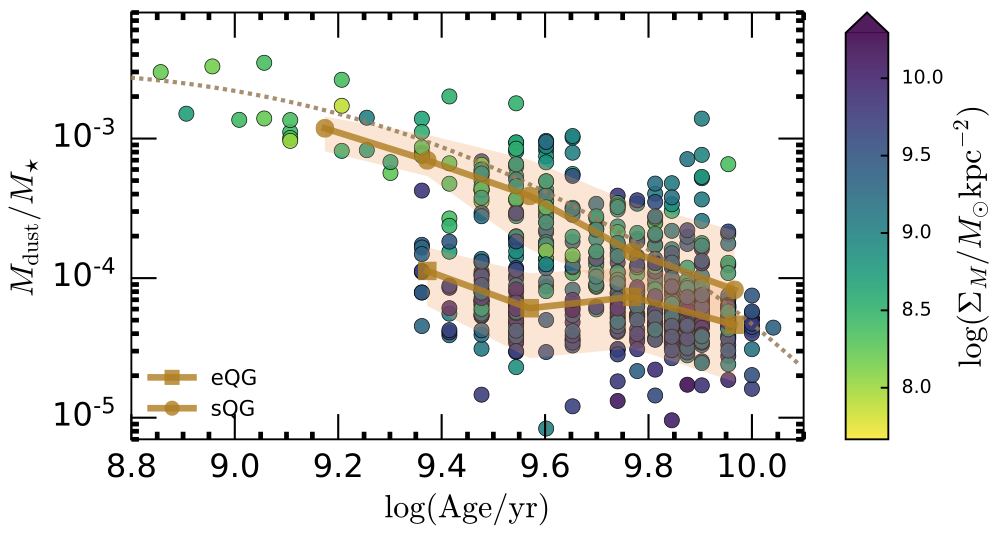}

		\caption{Evolution of $M_{\rm dust}/M_{\star}$ as a function of mass-weighted stellar age, colour-coded for the stellar-mass surface density. Brown symbols have the same meaning as in the previous Figure. Dotted line is the best fit (exponentially declining function) to the sQGs. 
		}
		
		\label{fig:Fig.5}
	\end{figure}

	We now focus on analysing the $M_{\rm dust}/M_{\star}$ under the age-evolutionary framework. To achieve this goal, we adopt the mass-weighted stellar ages ($t_{\star}$), derived from our self-consistent SED fitting. In \hyperref[fig:Fig.5]{Fig. 6} we show the trends of $M_{\rm dust}/M_{\star}$ with the mass-weighted stellar ages ($t_{\star}$) for two morphological subgroups. From \hyperref[fig:Fig.5]{Fig. 6} we see that $M_{\rm dust}/M_{\star}$ in sQGs has the overall decreasing trend with increasing stellar age, opposite to eQGs that maintain very shallow evolution. The median stellar age of sQGs is, on average, $\sim 1$ Gyr younger than in eQGs ($\log (\rm Age/yr)=9.65$ and $\log (\rm Age/yr)=9.79$, respectively). The large scatter is especially pronounced at older ages (e.g. $t_{\star}>5$ Gyr), while the difference between the two sub-groups almost vanishes towards the largest stellar ages. 
	
	Inferred mass-weighted stellar ages inform that the majority of stellar content was produced in early times in both morphological groups. There is a tail towards younger ages visible for the sQGs, while the same in not observed for the eQGs. The large spread in $t_{\star}$ as seen for our sQGs is in agreement with \cite{tojeiro13} and \cite{zhou22}, who found that some fraction of sQGs at low-$z$ have exceptionally younger ages (<0.5 Gyr) than the general population of eQGa and sQGs. By investigating the mass-size growth in the full sample of hCOSMOS QGs, \cite{damjanov22} show that the non-negligible number of extended QGs with younger stellar populations (as seen in our dustiest sQGs) are recently quenched galaxies. Interestingly, they found the evolution of the mass-size relation for those recently quenched QGs to be mostly a consequence of progenitor bias rather than of minor mergers.
	

	Negative correlation between $M_{\rm dust}/M_{\star}$ and stellar age (as seen in our sQGs) can be interpreted as age-evolutionary sequence (\citealt{michalowski19}, \citealt{burgarella20}). Within this framework, the level of specific dust decline with $\tau_{\star}$ can be used for quantifying the removal time for dust. For the small sample of H-ATLAS red spirals and ellipticals at $z\sim0.1$, \cite{michalowski19} found the exponentially declining function to the $M_{\rm dust}/M_{\star}$ with age, inferring long dust removal timescales ($\sim 2.25$ Gyr). If we apply the same method to our sQGs, we would obtain $\tau_{\rm d}=2.75\pm0.5$ Gyr, which agrees within $1\sigma$ with estimates from \cite{michalowski19}, despite some noticeable differences between the two samples.\footnote{The sample from \cite{michalowski19}  is composed of H-ATLAS objects that are described to have very bright SPIRE fluxes. In addition, their sample also contains "bluer" objects with $D_{\rm n}4000$ smaller than the cut we applied to systematically select our QGs ($D_{\rm n}4000$>1.5).} Contrary to sQGs, the evolution of $M_{\rm dust}/M_{\star}$ in eQGs is almost independent of stellar age, indicative of very slow dust removal, which likely be prolonged over several Gyr. Weak evolution of $M_{\rm dust}/M_{\star}$ at high stellar ages, and especially in eQGs, can strengthen the idea presented in previous Section that dust abundance in our QGs could be of mixed origin. For example, flattening of specific dust mass with age is expected if dust is continuously produced by AGB stars or is supported externally via mergers (see e.g., \citealt{smith14}, \citealt{richtler18}, \citealt{kokusho19}). An additional clue on this can be gained from mass surface densities. Many older QGs are generally expected to have undergone the size growth via minor mergers, which would increase the $M_{\star}$, but decrease the surface density. This would introduce the scatter in $\Sigma_{\rm M}$ per given age, in line to what is seen at \hyperref[fig:Fig.5]{Fig. 6}. Optical studies of QGs at low and intermediate redshifts (i.e., \citealt{beverage21}, \citealt{barone22}) pointed out that the scatter in surface densities per given age reflects the diverse conditions of the Universe when a galaxy becomes quiescent. 
	
	\subsection{The evolution of $M_{\rm dust}/M_{\star}$ with $M_{\star}$}
	\label{sec:5.3}
	
\begin{figure*}[h]
		\vspace{-0.2cm}
		\centering
		\includegraphics [width=14.99cm]{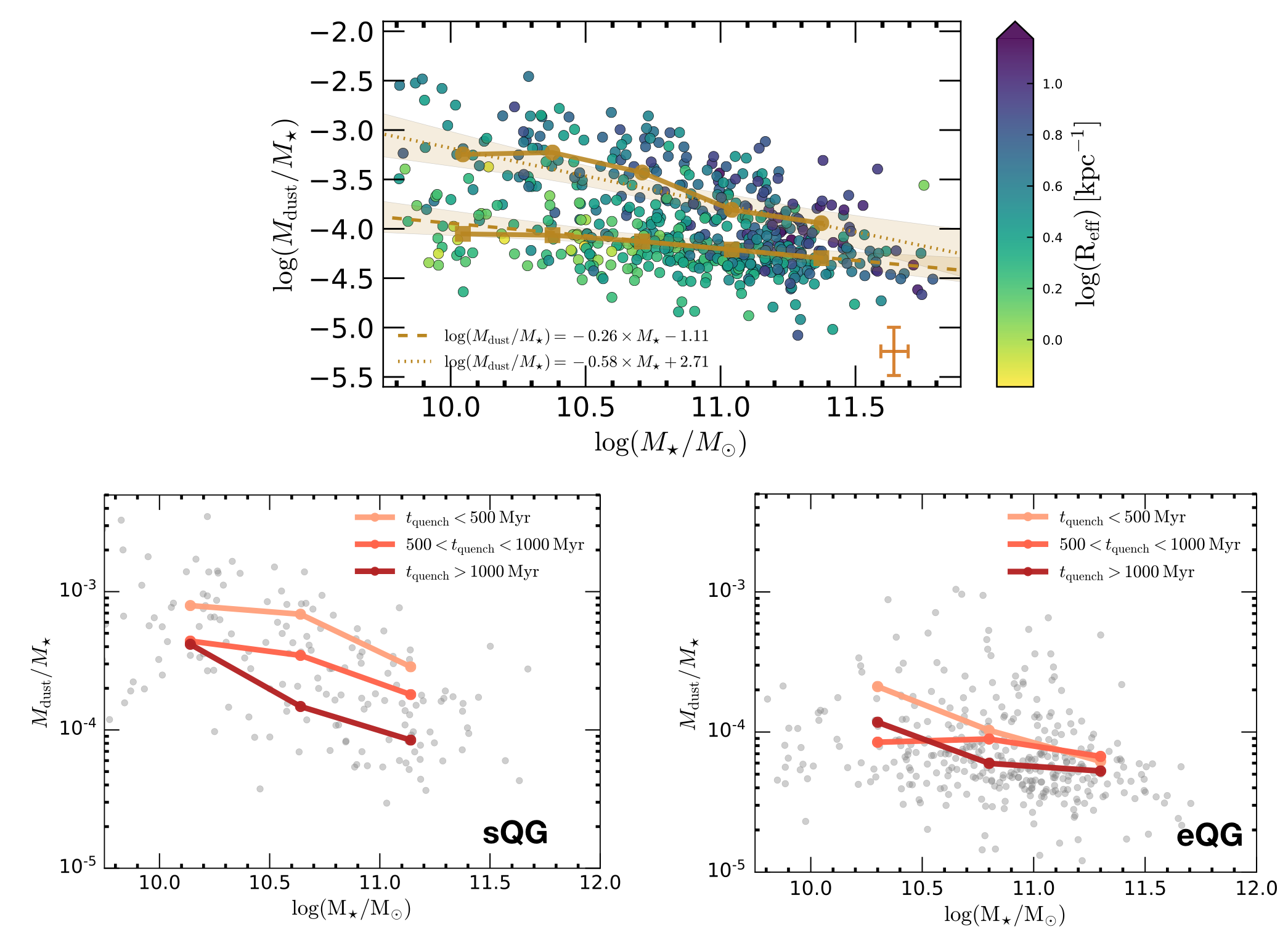}
		
		\caption{\texttt{Upper panel:} The evolution of $M_{\rm dust}/M_{\star}$ with $M_{\star}$ in our QGs, colour-coded as a function of galaxy size (represented with effective radius, $R_{\rm eff}$, in kpc). Median values in different stellar mass bins are displayed with brown circles and squares for sQGs and eQGs, respectively. The typical errors are shown as brown crosses. The straight lines are the best linear fits describing the data for sQGs and eQGs. Line symbols and fitting functions are indicated in the legend, while the shaded regions represent $95\%$ confidence interval of the fit; \texttt{Lower panel:} The evolution of $M_{\rm dust}/M_{\star}$ with $M_{\star}$ in sQGs (left) and eQGs (right), in the connection with time after quenching ($t_{\rm quench}$). Coloured lines are describing evolution of specific dust mass (as median values in three stellar mass bins) for different post-quenching intervals. The meaning of coloured lines is indicated in the legend.} 
		\label{fig:Fig.6}
	\end{figure*}

	Our data indicate a general anti-correlation between the $M_{\rm dust}/M_{\star}$ with $M_{\star}$. As evident from \hyperref[fig:Fig.6]{Fig.7}, for given $M_{\star}$, there is a distinction between the sQGs and eQGs at $\log(M_{\star}/M_{\odot})\lesssim 10.75$, with very little overlap between the two. Strong difference vanishes above this stellar mass with many sQGs entering the region populated by eQGs ($M_{\rm dust}/M_{\star}<10^{-4}$). The anti-correlation with $M_{\star}$ in sQG and eQGs can be best described with following fitting functions: 
	
	\begin{equation}
	\log(M_{\rm dust}/M_{\star})= -0.58\times M_{\star}+2.71
	\end{equation}
	for sQGs, and
	\begin{equation}
	\log(M_{\rm dust}/M_{\star})= -0.26\times M_{\star}-1.11
	\end{equation}
	for eQGs. The Spearman rank coefficient is find to be -0.62 for sQGs, and -0.32 for eQGs, while for both morphological groups we obtain a small probability (corresponding to the high significance of $\sim 5.5\sigma$ and $4.2\sigma$, respectively) that there is no correlation. Recent works on local dusty galaxies reported morphological impact on the relation between $M_{\rm dust}/M_{\star}$ with $M_{\star}$ (\citealt{nersesian19}, \citealt{casasola20}). Interestingly, the evolutionary slope we infer for our eQGs is identical to the one found in studies of nearby SF galaxies (\citealt{casasola20}), and in ALMA detected MS DSFGs at $z\sim 1$ (\citealt{donevski20}). This suggests that DSFGs and eQGs likely share some common processes in their dust evolution, i.e., removal and/or quenching (\citealt{lapi18}). 
	
	Negative correlation of $M_{\rm dust}/M_{\star}$ with $M_{\star}$ observed in our QGs is in agreement with known observational studies of dusty galaxies in different environments (e.g., \citealt{bourne12}, \citealt{casasola20}, \citealt{donevski20}). It is usually interpreted as a natural reflection of the dust life-cycle: $M_{\star}$ grows with time as galaxy evolve, while dust grains decrease from the budget being incorporated into the stellar mass.  By applying the chemical galaxy model on proto-spheroidal galaxies, \cite{calura17} find that $M_{\rm dust}/M_{\star}$ is larger in galaxies with $10<\log(M_{\star}/M_{\odot})<11$ because these are characterised both by a lower specific destruction rate and by a larger growth rate than in more massive galaxies. This may explain why our QGs with relatively high specific dust mass ($\log (M_{\rm dust}/M_{\star})>-3$) are identified exclusively at $\log(M_{\star}/M_{\odot})<11)$. \cite{casasola22} recently analyse the spatially resolved Dustpedia galaxies and show that variations in $M_{\rm dust}/M_{\star}$ exist even for galaxies within the same morphological group, attributing it to the variations in dust growth efficiencies. 
	
	There is another important aspect of our analysis presented in \hyperref[fig:Fig.6]{Fig.7}. Namely, 
	the uppermost part of the plane at $\log(M_{\star}/M_{\odot})\lesssim 11$ is dominated by sQGs that are characterised with larger sizes and stellar populations younger than in eQGs with same $M_{\star}$.  
	Further increase in optical sizes towards $\log(M_{\star}/M_{\odot})\gtrsim 11$ is followed by reduced difference between the specific dust masses of eQGs and sQGs. That said, the growth in galaxy size with $M_{\star}$ (as confirmed by many works on the mass-size relation) does not correspond to the similar increase in $M_{\rm dust}$. \cite{damjanov22} demonstrate that sizes of QGs from this work increase as $\log(R_{\rm eff}/\rm kpc)\sim0.75\times \log(M_{\star}/M_{\odot})$. While this evolution is find to be almost redshift independent, it emerges from different processes, both internal and external. Such dichotomy can modulate gas fraction and quenching efficiency which in turn would impact evolution of a dust content in galaxies. 
	
	
	\subsection {How the dust evolution and post-quenching timescales are connected? }
	\label{sec:5.4}
	
	One of the main questions that arise from presented results is: \textit{How can we use the observed evolution of dust content to learn about evolution of QGs following their quenching?} To answer this question, from star-formation histories modelled in CIGALE, we adopt post-quenching time ($t_{\rm quench}$), which is the time passed since the star formation is quenched, and is defined as age of the galaxy minus $t_{\rm trunc}$ from \hyperref[Eq.1]{Eq.1}. Our modelling procedure reveals wide distribution of $t_{\rm quench}$, from 60 Myr to 3.2 Gyr, which ensures that QGs from this work probe broad dynamical ranges in terms of quenching age. The median $t_{\rm quench}$ is found to be $436\pm 215$ Myr for sQGs, and $704\pm 282$ Myr for eQGs. Lower panel in \hyperref[fig:Fig.6]{Fig. 7} shows that QGs generally experience trend of $M_{\rm dust}/M_{\star}$ declining with post-quenching time increasing, which applies for both morphological groups. Morphology impact on $M_{\rm dust}/M_{\star}$ is evident by comparing the two sub-populations within the same range of $M_{\star}$ and $t_{\rm quench}$. The difference in $M_{\rm dust}/M_{\star}$ between the sQGs and eQGs is the largest in "recently-quenched" QGs ($t_{\rm quench}<500$ Myr), and smallest for "early-quenched" galaxies ($t_{\rm quench}>\rm 1Gyr$). Nevertheless, the average vertical drop in specific dust mass per given $M_{\star}$ is relatively small within the morphological groups: it is $\sim 0.5\:\rm dex$ for sQGs and even lower ($\sim 0.25\:\rm dex$) for eQGs. Therefore, the observed trend of $M_{\rm dust}/M_{\star}$ with $M_{\star}$ is mainly driven by increase of $M_{\star}$ with $t_{\rm quench}$, while $M_{\rm dust}$ exhibits remarkably shallow decline within the first $\sim1$ Gyr following quenching.
	
	Considering above, we can draw several conclusions: (1) Our data provide strong evidence for morphology playing a role in observed $M_{\rm dust}/M_{\star}$. Under the assumption that the level of SFR cessation correlates with $t_{\rm quench}$, we can deduce that for the majority of our QGs the removal of cold gas and dust are interlinked. However, our data strongly suggest the link is non-monotonic, which is manifested as flattening of trend between  $M_{\rm dust}/M_{\star}$ and $M_{\star}$ within certain $t_{\rm quench}$; (2) Such shallow evolution in $M_{\rm dust}/M_{\star}$ with increasing $M_{\star}$ cannot be explained with age-evolutionary scenario, and is rather indicative of \textit{"dust abundance recovery"} counterbalancing the drop of star-formation over different timescales. The "dust abundance recovery" in sQGs happens on timescales shorter than in eQGs ($t_{\rm quench}<500$ Myr vs. $t_{\rm quench}>500$ Myr, respectively). The difference may be caused by initial cold gas available after quenching which would affect the accretion timescales (see \hyperref[sec:5.5]{Section 5.5}). Evidence of flattening can be seen even in the most massive QGs that quenched $>1Gyr$ ago, but these objects have lower than average $M_{\rm dust}/M_{\star}$, which is likely a consequence of ISM removal mechanisms (e.g. gas heating, \citealt{smercina18}) starting to be visible against the prolonged dust production. 
	
	Dusty QGs from this work may serve as a benchmark for better understanding the connection between their structural properties and the ISM evolution. For instance, if $M_{\rm dust}/M_{\star}$ mirrors the change of molecular gas fraction (e.g., \citealt{imara19}, \citealt{whitaker21a}, \citealt{magdis21}), results from \hyperref[fig:Fig.6]{Fig. 7} implies that $f_{\rm gas}$ is the highest soon after the quenching episode, and that sQGs have on average a higher cold gas content than eQGs. 
	Within both morphological categories, more extended galaxies have larger specific dust mass. \cite{chen20} and \cite{barone22} argue that extended QGs with smaller surface densities undergo less extreme kinetic mode feedback (and consequently slower quenching), than similarly massive compact sources, which would leave them with more residual molecular gas. 
	
	
	Furthermore, the effective presence of dust in our QGs along with inferred $t_{\rm quench}$ may provide important clues on quenching mechanisms. As evolution in $M_{\rm dust}/M_{\star}$ changes its pace (\hyperref[fig:Fig.6]{Fig. 7} reveals phases that are both accelerated and slowed down within the same range of $t_{\rm quench}$), it is likely no one mechanism is responsible for quenching star formation in our QGs. This is consistent with observational studies of passive spirals (\citealt{fraser18}) and passive ellipticals (\citealt{zhou22}). It is also consistent with the TNG-50 simulation's study which reported variety of quenching timescales and time since quiescence, ranging from 1-4 Gyr in eQGs and sQGs (\citealt{park22}). 
	Similarly, \cite{mahajan20} found that local dusty QGs from GAMA survey are primarily affected by quenching acting on longer timescales (order of several Gyr) due to a lack of morphological transformation associated with the transition in optical colour. \cite{ciesla21} demonstrate how variations in dust properties (i.e. $L_{\rm dust}$) per given $t_{\rm quench}$ resemble the "strength" of quenching processes in a small sample of local dusty QGs from the \textit{Herschel} reference survey (HRS). They proposed how the dust abundance observed in objects that quenched $\sim 1$ Gyr ago can be explained if quenching is caused by ram-pressure stripping, which is known for affecting mostly gas, but not stars and dust in the central regions. The same mechanism is recently found to be responsible for galaxy size increase (\citealt{grishin21}) and some of our most extended sQGs with the highest specific dust masses appear to be compatible with this scenario. Interestingly, the average $M_{\rm dust}/M_{\star}$ as well as the large spread with stellar age, are comparable to those of dusty post-starbursts for which \citealt{lipsb19} reported offset of several hundred Myr between the SFR quenching and decline in gas-fraction and dust mass. 
	
	\subsection{Implications for the production of dust in QGs}
	\label{sec:5.5}
	Results presented throughout the \hyperref[sec:5]{Section 5} impose important constraints on the timelines and mechanisms of dust production and dust removal in QGs. Below we summarise possible scenarios, and continue with analysis with models in \hyperref[sec:6]{Section 6}.
	
	\textit{\textbf{Implications for internal dust growth:}} In galaxies dominated by older stellar populations, there will be time for AGB stars to contribute significantly to the dust mass budget. The total mass of dust formed by AGB intermediate mass stars ($1\:M_{\odot} \lesssim M_{\star}<8\:M_{\odot}$) depends on the initial mass of the star, with the maximal yield proposed to lay within $\sim 4\times 10^{-4}$ and $\sim 3\times 10^{-3}\:M_{\odot}$ for the case of solar and super-solar metallicities, respectively (\citealt{zhukovska08}, \citealt{ventura20}). By imposing maximal production in $\sim10^{10}$ AGB stars, we would obtain $M_{\rm dust}$ between $4\times 10^{6}M_{\odot}$ and $3\times 10^{7}M_{\odot}$. However, the average production efficiency by the AGB stars is at least 10-20 times (depending on the stellar model)  lower than the maximal one (\citealt{schneider14}), resulting in dust masses $\lesssim3\times 10^{6}M_{\odot}$. This is still below the average value of eQGs, suggesting insufficient dust yield solely produced by AGB stars to fully match our data.\footnote{It is worth noting here that this does not rule out contribution from AGB stars. Under certain conditions, they may be important suppliers of prolonged dust production (especially in eQGs), but it is unlikely for them to be the \textit{dominant} dust production channel.} On the other hand, average value in sQGs is even higher ($M_{\rm dust}>2.6\times10^{7}M_{\odot}$), difficult to explain even with maximally efficient SNe dust production in the early stages of galaxy history, as demonstrated by several works (i.e., \citealt{michalowski19}). Instead, our data favour dust grain growth as a relatively fast production mechanism to overcome the dust loss. As mentioned in \hyperref[sec:5.1]{Section 5.1}, in metal-rich galaxies this process can increase $M_{\rm dust}$ until metals are depleted from the ISM. Timescales for dust growth in QGs are proposed to be relatively rapid, up to $\sim 50-250$ Myr, if the lifetime of the cold gas is long enough ($\sim 10^7$ yr, \citealt{hirashita15}). If we stay conservative and account for the factor of $\sim3$ uncertainty due to expected variations in the $f_{\rm gas}$ of our QGs,  accretion timescales would remain shorter than $t_{\rm quench}$ for most of our galaxies. This may explain why the dust emission in our QGs is detectable in timescales longer than those expected for dust destruction due to gas heating ($\sim 10^8$ yr). 
	
	\textit{\textbf{Implications for external sources of dust:}} The most massive of our eQGs lack the clear evolution of $M_{\rm dust}/M_{\star}$ with stellar ages, metallicity and stellar mass. This qualitatively agrees with scenarios proposing external contribution to $M_{\rm dust}$ from minor mergers (\citealt{kartaltepe10}, \citealt{rowlands12}, \citealt{martini13}, \citealt{lianou16}), \citealt{richtler18}, \citealt{kokusho19}). Our sample contains only 17 QGs ($4\%$) with "disturbance" signatures (extended objects with morphologies that are on the edge between disks and irregulars). However, this should be considered a lower limit since there is a possibility that mergers occur long time ago enough to erase the clear disturbance signature. We thus follow the approach by \cite{nevin19} who investigate simulated mergers (both gas-poor and gas-rich) within a wide range of mass ratios. They leverage the strength of concentration coefficient and Gini index as the most sensitive predictors for minor mergers in their later stages, since these parameters generally enhance as the merger progresses. If we adopt the same criteria proposed by \cite{nevin19} (their Fig. 5), we would end up with 86 merger candidates or $16\%$ of our sample. Such criteria correspond to galaxies in their final, post-coalescence stage, $t>3.5$ Gyr since the merging started. Interestingly, the same authors find that the difference in $f_{\rm gas}$ between gas-poor and gas-rich mergers does not produce a significant distinction in the concentration of the remnant. Thus, we caution that candidate merging QGs in our sample can be a mix of both, gas-rich and gas-poor. In principle, galaxies that show strong positive excess in their $M_{\rm dust}/M_{\star}$ are likely to be compatible with the gas-rich scenario. The majority of our minor merger candidates (75/86) are eQGs of older ages ($>7-10$ Gyr) and higher stellar masses ($\log(M_{\star}/M_{\odot})>10.9$). Only 11/86 candidates are sQGs. As a result, we anticipate that internal mechanisms will dominate dust content in QGs, with minor mergers contributing more to eQGs (75/386, or $\sim19\%$) than sQGs (11/162, or $6\%$) We check the catalogue of merging \textit{Spitzer} galaxies in the COSMOS field (\citealt{kartaltepe10}) and find 14 objects from our sample that are present in this catalogue, and are classified as minor mergers. These QGs have $M_{\rm dust}/M_{\star}$ fluctuating within $\pm 0.3$ dex of the median of our full sample. In addition, there is a tentative evidence in our data that candidate mergers have $Z_{\rm gas}$ $\sim 0.1$ dex below the median, consistent with suggestions from the literature of mergers causing the scatter in mass-metallicity relation (\citealt{bustamante18}, \citealt{griffith19}). This all support the idea that merging environment can have visible impact on the evolution of $M_{\rm dust}/M_{\star}$ with $M_{\star}$, $Z_{\rm gas}$ and stellar age in QGs. Recent observational tests conducted with ALMA supported such possibility through identifying ISM complexity in a small sample of the brightest dusty eQGs at $z\sim0.05$ (\citealt{sansom19}). This study reveals examples of massive molecular gas discs, but also objects with dusty companions contributing to $>50\%$ of the FIR emission.
	\subsection {Implications for the survival of dust in massive QGs }
	\label{sec:5.6}
	The lack of evolution in specific dust mass of eQGs with $Z_{\rm gas}$, age and $M_{\star}$ is compatible with very slow dust removal. This can be a result of weak outflows, long destruction timescales due to SN shocks and inefficient thermal sputtering. Below we briefly examine some of the possibilities.
	\begin{itemize}
		\item \textit{Thermal sputtering}\\
		Galaxies in massive halos and clusters are expected to undergo significant thermal destruction of their grains. As presented in Donevski et a. (in prep), $\sim 15\%$ of QGs from this work reside in galaxy clusters. The observed $M_{\rm dust}/M_{\star}$ in these sources would be difficult to interpret if the sputtering timescale due to the exposition of dust grains to a hot ISM is extremely short. Timescales  within which dust grains are efficiently destroyed in QGs are given as $10^{5}(1+(10^{6} K/T_{\rm gas})^3)n_{e}^{-1}$ yr for typical 0.1 $\mu$m-sized grains (\citealt{hirashita17}). While the typical values are quite short (1-10 Myr, \citealt{vogelsberger19}), it would be possible to obtain longer destruction timescales ($t_{\rm sput}>1$ Gyr) if grains are larger and/or gas is of lower density (\citealt{smercina18}). Such longer sputtering timescales of $t_{\rm quench}>0.5-1$Gyr are another indication for prolonged dust growth as it supports the production of large grains which are expected to be reduced relatively slowly in hot halos (\citealt{relano22}, \citealt{priestley22}). On top of this, along with the sputtering in hot halos, dust cooling may be operating through the emission of IR radiation, as proposed by recent hydrodynamical simulations (\citealt{vogelsberger19}). However, this solution would likely produce excess in the sub-mm wavelengths, which we did not observe in existing SCUBA-2 data.\\
		
		\item  \textit{Destruction due to SN shocks}\\
		The total rate of dust mass destruction due to SN shocks is given by $\dot{M}_{\rm dest}\propto M_{\rm dust}/\tau_{\rm destr}$, where dust destruction timescale is usually approximated as (\citealt{slavin15}):
		
		\begin{equation}
		\tau_{\rm destr} = \frac{\Sigma M_{\rm gas}}{f_{\rm ISM}R_{\rm SN}M_{\rm cl}} = \frac{\tau_{\rm SN}M_{\rm gas}}{M_{\rm cl}}
		\end{equation}
		
		Here $\Sigma M_{\rm gas}$ is the surface density of molecular gas mass, $f_{\rm ISM}$ is the value that accounts for the effects of correlated SNe, $R_{\rm SN}$ is the SNe rate, $\tau_{\rm SN}$ is the mean interval between supernovae in the Galaxy (the inverse of the rate) and $M_{\rm cl}$ is the total ISM mass swept-up by a SN event. Here is important to note that $M_{\rm cl}$ varies with the ambient gas density and metallicity, and as metals being efficient cooling channel in the ISM, higher $Z_{\rm gas}$ would result in smaller swept mass (see \citealt{asano13}, \citealt{hou19}). We can roughly approximate range of destraction timescales by using estimated $Z_{\rm gas}$, and applying conservative approach for converting $M_{\rm dust}$ to $M_{\rm gas}$ through metallicity-dependent $\delta_{\mathrm{DGR}}$ (\citealt{remyruyer14}). Following Eq 4, we infer range of $\tau_{\rm destr}$ from $0.11\:\rm Gyr$ to $9.5\:\rm Gyr$, with the median of $1.28\:\rm Gyr$ for sQGs, and $1.75\:\rm Gyr$ for eQGs. This informs that destruction due to SNe shocks in not expected to be crucial in our sample. 
		
	\end{itemize}
Overall, the brief analysis from this section hints at an expected small influence of dust destruction due to SNe shocks, while effect of thermal sputtering can be significantly prolonged if grains are growing inside QGs. On top of this, effect of outflows (e.g., due to stellar and X-ray feedback) is expected to be more important (i.e., \citealt{delooze16}). To investigate this, in the next Section we analyse outputs from different simulations of dust formation that have these effects included.
	
	\section{Interpretation of results with models}
	\label{sec:6}

	To evaluate our results within the framework of dusty galaxy formation, in this section we inspect analytic models and the state-of-the-art cosmological simulations that track the dust life cycle in a self-consistent way.

	\subsection{Models}
	\label{sec:6.1}
	
	We analyse the predictions from the: \textbf{(I)} cosmological galaxy formation simulation with self-consistent dust growth and feedback (SIMBA, \citealt{simba}); and \textbf{(II)} chemical model of \cite{nanni20}.

	\subsubsection{SIMBA cosmological simulation}
	
	The cosmological galaxy formation simulation, \texttt{SIMBA}, utilizes mesh-free finite mass hydrodynamics, and is successor of MUFASA simulation (\citealt{mufasa}) with improvements to the sub-grid prescriptions for star formation, AGN feedback and X-ray feedback, as well as implementation of dust physics. We refer the reader to \cite{simba} for extensive description of the simulation. 
Self-consistent dust framework that models the production, growth and destruction of dust grains in \texttt{SIMBA} is introduced in \cite{li19}. Overall, \texttt{SIMBA} accounts for dust grains produced from stellar populations (SNe and AGB stars), which can further grow via accreting gas-phase metals. Dust can be destroyed via different processes such as thermal sputtering, SNe shocks and astration (consumption by star-formation).  Dust can instantaneously be destroyed in
gas impacted by jet-mode AGN feedback, while outflow processes such as radiative-mode Eddington AGN feedback and stellar feedback can heat up and transport dust out of the galaxy.

	
The net rate of dust production/destruction in \texttt{SIMBA} can be generalised as:
	\begin{equation}
	\label{eqn:4}
	\Sigma \dot{M}_{\rm dust}\propto \dot{M}_{\rm dust}^{\rm SNe}+\dot{M}_{\rm dust}^{\rm ISM}
	- \dot{M}_{\rm dust}^{\rm destr}- \dot{M}_{\rm dust}^{\rm SF}+ \dot{M}_{\rm dust}^{\rm inf}- \dot{M}_{\rm dust}^{\rm out}
	.\end{equation}
	
	The first term on the right side of \hyperref[eqn:4]{Eq. 7} describes the dust produced by condensation of a fraction of metals from the ejecta of $\rm SNe$ and $\rm AGB$ stars; the second term describes the dust by accretion in the ISM; the third term describes the dust destructed by $\rm SNe$ shock waves and thermal sputtering; the fourth term is the consumption of dust due to astration; the fifth term is an additional dust production by gas infall; the sixth term describes the dust mass that is ejected out of the ISM due to feedback processes. The latter two mechanisms are responsible for heating up and removal of gas into the DM halo, or even further out. The dust model within \texttt{SIMBA} does not include contribution from $\rm Ia \:SNe$ which is opposite to some models that proposed the same condensation efficiency between Type $\rm Ia\:SNe$ and Type II SNe (\citealt{dwek98}, \citealt{mckinon17}, \citealt{popping17}). The condensation efficiencies for AGB stars and and $\rm CC\:SNe$ are updated based on the theoretical models of \cite{ferrarotti06} and \cite{bianchi07}. The evolution of dust grains in ISM is simulated with single-sized dust model where the grains are assumed to all have the same initial radius and density ($a=0.1\:\mu$m and $\sigma=2.4\:g/cm^{3}$, respectively). 
	\texttt{SIMBA} was run on a number of volumes with different resolutions. Here we use the largest one which has $1024^{3}$ dark matter particles and $1024^{3}$ gas elements in a cube of $140\:{\rm Mpc} h^{-1}$ side length. This simulation run includes the full feedback and dust physics, which is ideal for our goal of studying the statistical sample of dusty QGs. We note that this  \texttt{SIMBA} run results in dust mass functions and dust-to-metal ratios in a good agreement with observations at $z<1$ (\citealt{li19}), and is able to explain the observed sub-mm number counts up to $z\sim2$ (\citealt{lovell20}) and the dust content of QGs observed at $z\sim1-2$ (\citealt{williams21}).	
	
	
	\subsubsection{The \cite{nanni20} model}
	
	The chemical model of \citet[hereafter \texttt{N20}] {nanni20} introduces a new set of analytical solutions for the dust evolution in galaxies. The model assumes that dust formed in supernovae remnants, around evolved AGB stars and in the ISM is composed predominantly by silicates (olivine and pyroxene), amorphous carbon dust and metallic iron. Similar to \texttt{SIMBA}, Type Ia SNe are not considered important contributors to the dust enrichment. In \texttt{N20} different physical processes are considered for affecting the time evolution of $M_{\rm dust}$: astration of gas and metals due to formation of stars; dust destruction by SN shock waves that propagate in the ISM, and dust removal by galactic outflows. The model does not include thermal sputtering of dust in hot halos. The following theoretical yields are adopted: \cite{kobayashi06} for AGB stars, \cite{cristallo15} for Type II SNe, and \cite{iwamoto99} for Type Ia SNe Consistently with the calculations of gas and metal evolution, \texttt{N20} accounts for the inflow material being composed of pristine gas that does not change the total content of dust. At each time-step the galactic outflow is considered to be regulated through the "mass-loading factor" (ML) and assumes the outflow to be due to the stellar feedback in the ISM. The ML parameter is quantitatively defined as the ratio of outflow rate and SFR ($\rm ML=\dot{M}_{\rm out}/\rm SFR$) and is related to initial baryon mass in the galaxy, that is fully composed of gas at the beginning of simulation and converted to stars afterwords. In \texttt{N20} model, $M_{\rm gas}$ is set as a multiple of $M_{\star}$ normalised to 1$\:M_{\odot}$ at 13 Gyr. While the fiducial \texttt{N20} model excludes explicit growth of dust particles in ISM, here we test the importance of this mechanism by switching it on and off in different simulation runs. It is worth noting that \texttt{N20} model showed success in reproducing total dust mass and star formation rate in nearby galaxies and their Lyman-Break galaxy (LBG) analogues at high-$z$ (\citealt{burgarella20}).
	

	\subsection{Comparison between models and data}
	
	\label{sec:6.2}
	We now confront our observational findings to the models described above. To ensure consistency between observed and simulated data, we impose the same range of modelled $M_{\star}$, $M_{\rm dust}$  and $\rm sSFR$ as in our observations ($\log (M_{\star}/M_{\odot})>10$, $\log (M_{\rm dust}/M_{\odot})>6$ and $\log(\rm sSFR/\rm yr^{-1})<-10$). In addition, we follow \cite{damjanov18} and infer the $90\%$ stellar-mass completeness limit of our hCOSMOS data by fitting the function $M_{\star}=10.6-\log(0.4/z)-0.2$. We find that 46 dusty QGs reside below the completeness limit defined in this way. Furthermore, only sources identified within the $90\%$ completeness limit are taken into account when comparing to models. We note that same IMF (\citealt{chabrier03}) is adopted both in simulations, and for the SED fitting of observed data. 
	
	\begin{figure*}
		\centering
		
		\includegraphics [width=17.69cm]{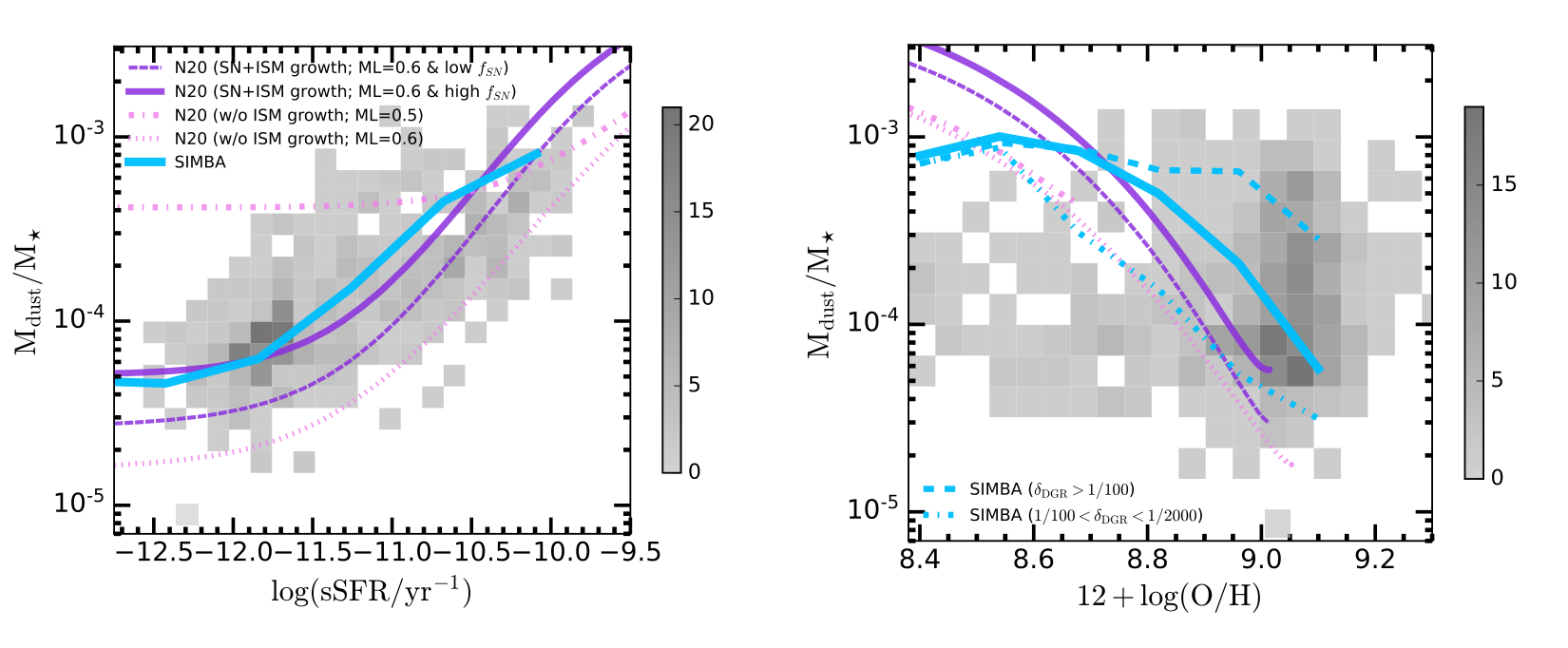}

		\caption{Evolution of $M_{\rm dust}$/$M_{\star}$ as a function of $\rm sSFR$ (\textit{left}) and $Z_{\rm gas}$ (\textit{right}) modelled with the state-of-the-art cosmological simulation SIMBA (\citealt{simba}, blue lines) and chemical evolution model of \cite{nanni20} (light and dark violet lines). In the panels, darker (lighter) curves related to \texttt{N20} model are simulations that include (exclude) dust grain growth in the ISM, respectively. These simulation runs include different values of mass-loading factors and the grain condensation efficiencies, as indicated in the legend. Solid blue line shows the SIMBA model prediction which is the result of the flagship run (100 Mpc/h box) that includes ISM dust growth on metals and all feedback variants. In the right panel, along with the median trend for all selected dusty QGs in SIMBA, we also display average tracks for QGs with different $\delta_{\rm DGR}$ (blue dashed line for $\delta_{\rm DGR}<1/100$, and blue dotted-dashed line for $1/100<\delta_{\rm DGR}<1/2000$). Data from this work (contributing to the $\sim90\%$ completeness) are displayed as binned 2D histograms, with colorbar showing the number of observed QGs in each cell.} 
		
		\label{fig:Fig.7}
	\end{figure*}
	
	In \hyperref[fig:Fig.7]{Fig. 8} we show how the $M_{\rm dust}/M_{\star}$ in models evolves as a function of sSFR and $Z_{\rm gas}$. Predictions from \texttt{SIMBA} match our data very well as the simulation shows a great potential of simultaneously reproducing observed trends of $M_{\rm dust}/M_{\star}$ with sSFR and $Z_{\rm gas}$. 
	
	For the \texttt{N20} model we showcase predictions of different simulation runs. These account for different outflow rates (ML=0.5, 0.55 and 0.6) and different dust condensation fractions by Type II SNe (low fraction of $25\%$ and high fraction of $50\%$). The total ISM mass swept in each SNe event is adopted to be $M_{\rm sw}=6800 M_{\odot}$, which is a standard value proposed in the literature (\citealp{dwek14}). We also explore cases where we switch on/off modelled dust growth in ISM. We find that the evolution of $M_{\rm dust}/M_{\star}$ with sSFR in \texttt{N20} is well reproduced when the dust grain growth in the ISM is included, and is accompanied with the moderate outflow and higher dust condensation fraction ($\sim 50\%$). By performing controlled simulation experiments with \texttt{N20} we find the evolution of specific dust mass is extremely sensitive to the changes of ML factor. Namely, the higher outflow rates ($\rm ML>0.6$) would rapidly reduce $M_{\rm dust}/M_{\star}$ to the values lower than $10^{-5}$ in older stellar populations ($>5-10$ Gyr). This would produce swift decline of specific dust mass with sSFR. In addition, for weaker outflows $M_{\rm dust}/M_{\star}$ remains too flat for decreasing values of the sSFR, indicating that dust astration and destruction from SNe are not sufficient to explain the anti-correlation of $M_{\rm dust}/M_{\star}$ with sSFR, even if we impose maximum efficiency of dust destruction. While the best-matching model of \texttt{N20} reproduces observed trend of $M_{\rm dust}/M_{\star}$ with sSFR, at the same time it falls relatively short to reproduce the relation with $Z_{\rm gas}$. As seen from \hyperref[fig:Fig.7]{Fig. 8}, $M_{\rm dust}/M_{\star}$ are 4-5 times too low for the values we infer in super-solar $Z_{\rm gas}$, while decline in dust abundance is slightly steeper as compared to \texttt{SIMBA}. At (sub-)solar $Z_{\rm gas}$ both models predict higher $M_{\rm dust}/M_{\star}$ than inferred by our data. As we note in \hyperref[sec:5.1]{Section 5.1}, the difference may be just an artefact due to large uncertainties on estimated $Z_{\rm gas}$. Otherwise, lower $M_{\rm dust}/M_{\star}$ than seen in models for (sub-)solar $Z_{\rm gas}$ may indicate interesting case of objects missed by simulations, e.g. galaxies that accrete gaseous material from the cosmic web, but still did not have enough time to significantly increase their dust abundance. 
	
	\subsection{Sources of dust in QGs at intermediate redshifts: insights from models}
	
	Confronting the different theoretical models and simulations allows us to better understand the contributors for dust abundance in QGs. Results from both observations and models presented throughout this paper strongly support the scenario where dust growth due to grain collisions in the metal-rich ISM may play a vital role in dusty QGs. This conclusion is in general agreement with other observational studies exploring dust content in QGs (\citealt{michalowski19}, \citealt{richtler18}), and with recent theoretical predictions (\citealt{hirashita15}, \citealt{hirashita17}, \citealt{vijayan19} and \citealt{triani21}). Indeed, it has been demonstrated that excluding ISM grain growth in \texttt{SIMBA} yields $M_{\rm dust}$ that are $\sim0.5$ dex lower than in QGs for which ISM dust growth is included (e.g., \citealt{whitaker21b}). Similarly, by turning off the effect of ISM grain growth in \texttt{N20}, simulated $M_{\rm dust}/M_{\star}$ are lower by $\gtrsim 0.4$ dex as compared to cases when dust growth is included.\footnote{Note that various combinations of galactic outflows and condensation efficiencies nay introduce some degeneracies in the models, making the predictions strongly dependent on how one regulates the outflow (see \citealt{nanni20} for the discussion.} We note that the shortfall to data in \texttt{N20} cannot be cured by imposing lower ML factors which would result in an overall flat trend with sSFR, inconsistent with our data (left panel of \hyperref[fig:Fig.7]{Fig. 6}). Interestingly, \cite{nanni20} found that dust growth has irrelevant role in the dust content in LBGs and local dwarf galaxies. This clearly demonstrates that different mechanisms for the dust build-up are in place in later phases of massive galaxy evolution. 
	
	Since the timescale for dust growth in the ISM changes as a function of gas surface density for given $Z_{\rm gas}$, good agreement of our results with \texttt{SIMBA} implies that dusty QGs in hCOSMOS have enough available cold gas to support dust growth over quite short timescales ($\sim 100-250$ Myr). \texttt{SIMBA} predicts median gas masses of $M_{\rm gas}\sim2\times10^9\:M_{\odot}$, but higher $f_{\rm gas}$ in sQGs than in eQGs ($f_{\rm gas}=0.11\pm 0.04$ vs. $f_{\rm gas}=0.06\pm0.03$ respectively). In addition, while checking simulation results, we also notice that sQGs in SIMBA tend to have more extended distributions of cold gas mass as compared to the stellar mass, which is not the case in eQGs. This can be explained either with higher ratios of accreting gas from the outskirts after minor mergers, or that gas is mostly distributed in the disc regions, while the stellar component is very compact. Such difference in $f_{\rm gas}$ support our earlier qualitative conclusion from \hyperlink{sec:5.4}{Section 5.4} and \hyperlink{sec:5.5}{Section 5.5}, and may be the at the heart of why observed $M_{\rm dust}/M_{\star}$ is larger in sQGs than in eQGs. 
	
	Quenching timescales in \texttt{SIMBA} are strongly bimodal, with the rapid ones ($<250-500$ Myr) being overly effective within very specific mass range ($10<\log(M_{\star}/M_{\odot})<10.5$) due to black hole feedback mechanism (\citealt{montero19}, \citealt{zheng22}). The vast majority of dusty QGs that quench at higher stellar masses ($\log(M_{\star}/M_{\odot})>10.5$) are expected to be preferentially influenced by the "slow quenching" mode (order of $\gtrsim 0.5-2$ Gyr). This mass range corresponds to the $\sim75\%$ of our sources. Observed shallower decline and large spread of $M_{\rm dust}/M_{\star}$ towards older ages and higher $M_{\star}$ are consistent with this scenario. While black hole feedback in \texttt{SIMBA} is responsible for heating halo gas, it explicitly does not affect most of the cold gas in the ISM since it is hydrodynamically decoupled. In other words, even if there is no star-formation, there still would be a material to support the dust growth within $\sim 10^7-10^8$ yr after the quenching. Consequently, prolonged growth of dust would not be possible in the case of X-ray feedback as it has been showed that in SIMBA it efficiently pushes central gas outwards over very short timescales (\citealt{appleby20}).
	
	Requirement that ISM growth is needed to simultaneously explain the evolution of specific dust mass with sSFR and $Z_{\rm gas}$ also supports predictions of semi-analytical model by \cite{triani21}. They show that lack of strong SNe destruction events in dust-rich quenched galaxies supports the super-solar $Z_{\rm gas}$ and prolonged build-up of dust. While abundance of metals in the ISM provides material for galaxies to accumulate dust mass, stating that grain growth is the only process responsible for observed $M_{\rm dust}$ in QGs can be misleading. In the previous section we offer different lines of evidences that $\sim 15\%$ of our QGs obey signs of (past) merging activities. Accretion of a gas-rich satellite would sustain prolonged dust grain growth in the accreted cold ISM if $Z_{\rm gas}$ is above certain threshold\footnote{The exact value of $Z_{\rm gas}$ where dust growth becomes dominant over stellar production is proposed to vary. For example, recent analytic models suggest $\rm 12+\log(O/H)>8.5$ for $M_{\rm gas}>10^9M_{\odot}$, e.g., \cite{triani20}}. We however expect that minor mergers play secondary role in the dust build up in dusty QGs in \texttt{SIMBA}. \cite{montero19} show that there is a small fraction ($<10\%$) of low-level rejuvenation events mostly related to minor mergers with gas-rich satellites at $z<1$. These processes contribute to the young discs and more cold gas in the outskirts, and are followed by fast ($\sim 100$ Myr) quenching events. 
	
	
	Aside from the two favourable mechanisms outlined above, there are other possibilities that might be responsible for the dust content observed in our QGs. \cite{nanni20} claims that using a non-standard "top-heavy" IMF increases dust yields by a factor of $\times\:3-4$ if a non-standard 'top-heavy' IMF is applied (but, see also discussion by \citealt{mckinon17}). At the same time, we expect a non trivial connection between dusty QGs and their large-scale environments. The detailed analysis of this interplay is out of the scope of this paper, but is addressed in our ongoing work (Donevski \& Kraljic, in prep.).
	
	\subsection{To which extent  does the specific dust mass mirror the molecular gas fraction in QGs?}
	\label{sec:6.4}
	Finally, relying on a good agreement with \texttt{SIMBA}, we link our observed and simulated data to inform future observations on expected interplay between the dust, gas and metals in QGs. In previous sections we show that observed $M_{\rm dust}/M_{\star}$ can be resulting from a dust growth, for which timescales may be offsetting from those attributed to the decline of star-formation and gas exhaustion. We can question: \textit{To which extent the observed evolution of $M_{\rm dust}/M_{\star}$ mirrors the evolution of molecular gas?} To give a sense of how the evolution of $M_{\rm dust}/M_{\star}$ with $Z_{\rm gas}$ can be used to deduce the cold gas-mass properties, we follow \cite{tan14} and rewrite the $M_{\rm dust}/M_{\star}$ as: 
	
	\begin{equation}
	\label{eqn:Eq.1}
	\frac{M_{\rm dust}}{M_{\star}}\propto\frac{M_{\rm gas}}{M_{\star}} \times Z_{\rm gas}\times \delta_{\mathrm{DTM}},
	\end{equation}
	
	where $M_{\rm gas}/M_{\star}$ is molecular gas ratio, and $\delta_{\mathrm{DTM}}$ is dust-to-metal ratio. 
	
	\begin{figure*}
		\centering
		
		\includegraphics [width=12.69cm]{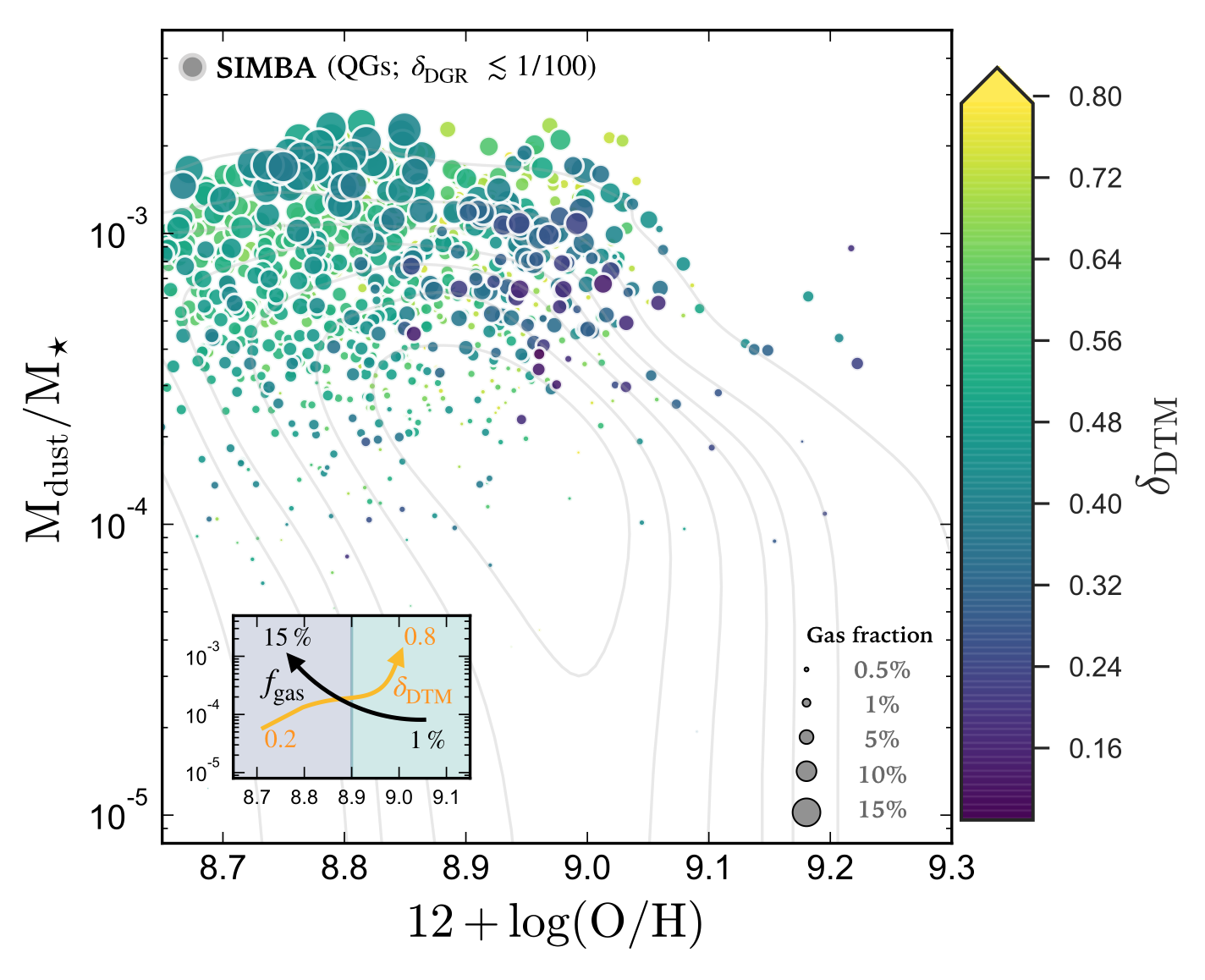}
		\includegraphics[width=10.09cm]{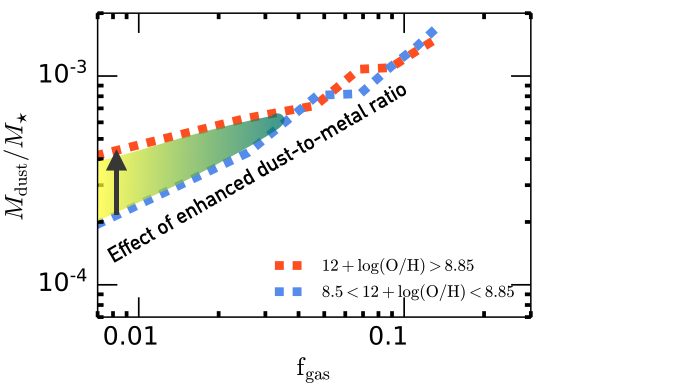}
		
		\caption{\texttt{Upper panel:} Modelled $M_{\rm dust}/M_{\star}$ as a function of $Z_{\rm gas}$ in \texttt{SIMBA}. Circles represent the modelled QGs within the restricted range of dust-to-gas mass ratios ($1/20<\delta_{\mathrm{DGR}}<1/100$). Grey contours reveal the distribution of all QGs selected in the simulations with the criteria described in \hyperref[sec:6]{Section 6.1}. The points are colour-coded for the $\delta_{\rm{DTM}}$, with circle sizes scaling to $\rm f_{\rm gas}$. Inset sketch qualitatively describes the behaviour of $M_{\rm dust}/M_{\star}$: up to the certain $Z_{\rm gas}$ it mostly resembles change of molecular gas fraction, while at $\rm 12+\log(O/H)\gtrsim 8.9$ it reflects the level of dust growth through the gradual increase in $\delta_{\rm{DTM}}$; bbcbbc\texttt{Lower panel:} Evolution of $M_{\rm dust}/M_{\star}$ with $f_{\rm gas}$ for the same simulated sources plotted in the upper panel. The blue and orange dashed lines show median trends for two regimes of $Z_{\rm gas}$ as indicated in the legend. The effect of increased dust growth in QGs is most pronounced in sources with heavily exhausted gas reservoirs, as indicated by lower molecular gas fractions ($f_{\rm gas}<1-5\%$). To highlight this effect through the rise in $\delta_{\rm{DTM}}$, we sketch the shaded area between the two tracks and colour-code it to roughly match the colorbar convention from the upper panel.
		}
		\label{fig:Fig.8}
	\end{figure*}
	
	There are two possible approaches to solve this equation. One is to follow the method applied by \cite{magdis21}, who assumed constant, metallicity dependent $\delta_{\mathrm{DGR}}$ for all QGs (they adopt $\delta_{\mathrm{DGR}}=1/92(1/35))$ for solar and super-solar $Z_{\rm gas}$, respectively). However, this solution implies a constant dust-to-metal ratio, in which case the evolution of $M_{\rm dust}/M_{\star}$ entirely mirrors the evolution of molecular gas fraction. While this offers an appealing empirical explanation for the observed evolution of the ISM properties of QGs, it does not account for the possible variations in dust-to-metal ratio. Indeed, $\delta_{\mathrm{DTM}}$ would remain constant if dust and metals are created from stars at same rates throughout galaxy evolution. Otherwise, with metals supporting the dust grain growth, $\delta_{\mathrm{DTM}}$ would change and becomes important marker of the growth efficiency (i.e., \citealt{feldman15}, \citealt{deVis19}). 
	
	We thus apply another approach which benefits from the good agreement with \texttt{SIMBA}. We explore what lies behind the scatter in $M_{\rm gas}/M_{\star}$ at a given $Z_{\rm gas}$ in simulations. To illustrate how $f_{\rm gas}$ and $\delta_{\mathrm{DTM}}$ relate to $M_{\rm dust}/M_{\star}$, in \hyperref[fig:Fig.8]{Fig.9} we display simulated QGs that have $\delta_{\mathrm{DGR}}$ within the restricted range of values that are usually adopted in observations to convert observed $M_{\rm dust}$ to $M_{\rm gas}$ ($1/20<\delta_{\mathrm{DGR}}<1/100$; e.g., \citealt{leroy11}, \citealt{schreiber18}, \citealt{as2uds19}, \citealt{whitaker21a}, \citealt{magdis21}). From the upper panel of \hyperref[fig:Fig.8]{Fig.9} we see that even for such a relatively small difference in $\delta_{\mathrm{DGR}}$, specific dust masses display wider spread in values ($\sim8\times 10^{-5}<M_{\rm dust}/M_{\star}<3\times 10^{-3}$).
	
	Most importantly, \hyperref[fig:Fig.8]{Fig.9} reveals that gas fraction is not the faithful tracer of $M_{\rm dust}/M_{\star}$ over the entire range of $Z_{\rm gas}$. \texttt{SIMBA} predicts that up to $\rm 12+\log(O/H)\lesssim8.85-8.9$, observed $M_{\rm dust}/M_{\star}$ mainly traces $f_{\rm gas}$, but this correlation is weaker for $12+\log(\rm O/H)\gtrsim 8.85-8.9$, when increase in $\delta_{\mathrm{DTM}}$ becomes more important. In other words, it is gradual evolution of $\delta_{\mathrm{DTM}}$, rather than of gas fraction, that are responsible for the spread in observed specific dust mass in super-solar dusty QGs. Therefore, we should expect that some of very metal-rich QGs with $\delta_{\mathrm{DTM}}>0.5$ may still have relatively high specific dust mass ($10^{-3}<M_{\rm dust}/M_{\star}<10^{-4}$) despite heavily exhausted gas reservoirs ($f_{\rm gas}<1-5\%$). This is illustrated in the lower panel of \hyperref[fig:Fig.8]{Fig.9}. The effect of prolonged dust growth on metals causes different pathways (slower and faster) for the decline in $M_{\rm dust}/M_{\star}$. The corresponding offset in $M_{\rm dust}/M_{\star}$ starts to increase below certain $f_{\rm gas}$. In galaxies that show positive excess, thermal sputtering acts on longer timescales ($>1$ Gyr) due to inefficiency in swiftly destroying larger grains produced via grain growth. If this is not the case, QGs would end up with extremely low $\delta_{\mathrm{DGR}}$ as reported in some dusty QGs observed in the local Universe (e.g., \citealt{lianou16}, \citealt{casasola20}). For that reason, any use of $M_{\rm dust}/M_{\star}$ to convert to $f_{\rm gas}$ and ultimately to $M_{\rm gas}$ must be carefully treated when QGs have super-solar $Z_{\rm gas}$.  This is particularly important for planning the observations (i.e., with ALMA) to detect the molecular gas via CO[2-1]. At a given $M_{\rm dust}/M_{\star}$, galaxies with the highest $Z_{\rm gas}$ would require significantly longer integration time than those with $Z_{\rm gas}\sim Z_{\odot}$ due to their higher $\delta_{\mathrm{DGR}}$ and subsequently lower $M_{\rm gas}$. This suggests that care should be taken when interpreting the nature of dust evolution in metal-rich QGs. 

\section{Conclusions}
\label{sec:7}

We present the first statistical study of the evolution of dust-related properties in QGs observed at intermediate redshifts ($0.1<z<0.6$) as part of the hCOSMOS spectroscopic survey. We analyse 548 massive ($M_{\star}>10^{10}\:M_{\odot}$) QGs that are systematically selected based on their spectral age indices ($D_{\rm n}4000>1.5$). We combine spectroscopic information with a self-consistent, multi-band SED fitting method to explore trends of $M_{\rm dust}/M_{\star}$ with various physical parameters, including independently measured gas-phase metallicity. This is the largest sample of QGs for which the evolution of dust, metal and stellar content have been estimated to date. We fully evaluate our findings with state-of-the-art simulations of dusty galaxy formation. Our main results are summarised as follows: 

\begin{itemize}
	\item Dusty QGs at intermediate redshifts experience the cosmic evolution in $M_{\rm dust}/M_{\star}$, but we find it to be shallower than in known studies on stacked samples. A large spread ($>2$ orders of magnitude) in $M_{\rm dust}$/$M_{\rm \star}$  suggests non-uniform ISM conditions in these massive galaxies. 
	
	
	\item Morphological and structural parameters are important factors in the scatter in specific dust masses. The $M_{\rm dust}/M_{\rm \star}$ in sQGs is $\gtrsim4$ times greater than in eQGs of a similar $M_{\star}$. At $M_{\star}<10^{11}M_{\odot}$, higher specific dust masses are associated with lower stellar mass surface densities and younger ages of stellar populations, whereas the same relationship is less obvious at $M_{\star}>10^{11}M_{\odot}$.
	
    \item The sQGs exhibit evolutionary trends of $M_{\rm dust}/M_{\rm \star}$ with $M_{\star}$, stellar age and galaxy size, in contrast to little to no evolution in eQGs. 
	
	\item Although both sQGs and eQGs have median super-solar $Z_{\rm gas}$ ($12+\log(\rm O/H)\sim 9$), sQGs have a reversal of $M_{\rm dust}/M_{\rm \star}$ towards the highest $Z_{\rm gas}$, whereas eQGs have a flat trend. We interpret this as a change in efficiency of dust production and removal since quenching. 
	
	\item We derive a broad dynamical range of post-quenching timescales in QGs ($60\:\rm Myr<t_{\rm quench}<3.2\:\rm Gyr$). In general, $M_{\rm dust}/M_{\rm \star}$ is the highest in recently quenched systems ($t_{\rm quench}<500$ Myr), but its further evolution is non-monotonic, which is strong evidence for different pathways for prolonged dust growth, or removal on various timescales. The moderately shallow evolution of $M_{\rm dust}/M_{\rm \star}$ imply that quenching mechanisms usually attributed to quick removal of gas and dust (i.e., powerful outflows and X-ray feedback) are not overly influential within our sample.
	
	
	
	
	\item State-of-the art cosmological simulation \texttt{SIMBA} and the chemical model of \cite{nanni20} are both capable of producing the dustiest QGs observed in this work ($M_{\rm dust}/M_{\star}\gtrsim10^{-3}$), with the difference that chemical models fall short of fully explaining the observed specific dust masses at super-solar $Z_{\rm gas}$. The prolonged grain growth on ISM metals is strongly required in simulations to account for the observed dust content in QGs. Without this assumption, the models have a larger discrepancy with the data, even when adopting substantial stellar yields with high condensation efficiencies and moderate outflows.
	
    \item Our results strongly suggest that the observed $M_{\rm dust}$/$M_{\rm \star}$ in dusty QGs is twofold: (1) The dominant channel is prolonged dust growth in metal-rich ISM. Such a process is viable in the first 1 Gyr since quenching, and it helps dusty QGs significantly prolong the destruction timescales expected for gas heating; (2) For $\sim 15\%$ of sources we find evidence that the dust content is supported or acquired externally, most likely via minor mergers.
	
	\item Cosmological simulation \texttt{SIMBA} predicts that enhanced $M_{\rm dust}/M_{\rm \star}$ in many QGs with super-solar $Z_{\rm gas}$ mirrors the enhancement in dust-to-metal ratio rather than rise in molecular gas fraction. This prediction has an important observational consequence in that QGs can still be observed as dust-rich ($M_{\rm dust}/M_{\rm \star}\simeq10^{-3}$) even if their gas reservoirs are heavily exhausted (with the gas fraction $<1-5\%$).

\end{itemize} 

	
The complexity of galaxies' dust life-cycles revealed by our study motivates future observational investigations of dusty QGs. 
The model predictions described in \hyperref[sec:6]{Section 6.} (in particular the diversity in $\delta_{\rm{DTM}}$) are rooted in the mechanisms dominating dust production and removal. Many observational tests of these predictions are now directly accessible via JWST instruments. 
The JWST MIRI, in particular, can be used for probing the warm dust and molecular gas through mid-IR diagnostics in individual dusty QGs at various cosmic epochs up to $z\sim2-3$. Along with independent metallicity measurements, the JWST MIRI can be used to place strong constraints on the fractions and sizes of PAH grains through multiple PAH features with a large wavelength separation. At the same time, it can help unveiling the impact of dust-hidden AGN activity that would influence the dust removal. Furthermore,  silicate strengths and a rising mid-IR emission longwards of $\geq 10\mu$m would offer insights on dust grain growth. There are numerous synergistic opportunities with high-resolution sub-mm observations (e.g., with ALMA and NOEMA) that would provide information on the cold dust and gas in QGs. Finally, in our ongoing works, we examine how the large-scale environments influence the interplay of dust and metals in QGs (Donevski \& Kraljic, in prep.) and present the key differences in SEDs of dusty vs. non-dusty QGs (Lorenzon et al., in prep.).

	\begin{acknowledgements}
We are thankful to the anonymous referee for very constructive comments which  improved the initial version of the paper. We would like to acknowledge Anna Feltre, Kasia Malek, Marcella Massardi, Lorenzo Zanisi, Katarina Kraljic and Alessandro Bressan for useful discussions and comments D.D acknowledges support from the National Science Center (NCN) grant SONATA (UMO-2020/39/D/ST9/00720). I.D. acknowledges the support of the Canada Research Chair Program and the Natural Sciences and Engineering Research Council of Canada (NSERC; funding reference No. RGPIN-2018-05425). A.M. acknowledges the support of the Natural Sciences and Engineering Research Council of Canada (NSERC) through grant reference number RGPIN-2021-03046. M.R. and A.N. acknowledge support from the Narodowe Centrum Nauki (UMO-2020/38/E/ST9/00077). D.N. acknowledges support from the NSF via AST-1909153. 

	\end{acknowledgements}
	
	
	\bibliographystyle{aa}
	\bibliography{ddrisers}
	
	
	\begin{appendix} 
		
				\section{Gas-phase metallicities}
				\label{App.A}
		
		We derive the gas phase metallicity of hCOSMOS galaxies following the method described in \citet{Zahid13} and \citet{Sohn19}, which is based on measuring the emission line strengths. We first derive the model continuum of each galaxy based on the stellar population synthesis model by \citet{bruzual03}. Then, we fit each emission line with a Gaussian from the continuum subtracted spectrum. The uncertainties in the line flux measurements were also computed using standard error propagation from the uncertainties in the spectrum. We then apply dust extinction correction based on \citet{Cardelli89} extinction curve.
		
		We apply the \citet{Kobulnicky04} technique that determines the gas metallicities (i.e., 12 + log (O/H)) of galaxies. This technique uses $R_{23}$ and $O_{32}$ indices that are defined as:
		\begin{equation}
		R_{23} = \frac{[\rm OII]\lambda3727 + [\rm OIII]\lambda4959 + [\rm OIII]\lambda5007 }{H\beta},
		\end{equation}
		and
		\begin{equation}
		O_{32} = \frac{[\rm OIII]\lambda4959 + [\rm OIII]\lambda5007}{[\rm OII]\lambda3727}.
		\end{equation}
		
		Here, we use the line fluxes of each emission line. We note that [OII]$\lambda3727$ indicates the sum [OII]$\lambda 3726 + \lambda 3729$ doublet, which is hardly resolved in Hectospec spectra. We also use 1.33 times [OIII]$\lambda 5007$ as the sum of [OIII] line flux based on the assumption that the flux ratio [OIII]$\lambda 4959$ / [OIII]$\lambda 5007$ is equal to 3 (\citealt{Osterbrock06}). We finally determine the metallicity based on the relative positions of $R_{23}$ and $O_{23}$ indices with respect to model grids with an intrinsic measurement uncertainty of $\sim 0.1$ dex \citep{Kobulnicky04}. All of our sources lack $H\alpha$ while vast majority of sources ($>85\%$) have $H\beta$ emission strength of $\sim0.2-0.3$ dex lower than the median of SF hCOSMOS galaxies, confirming their quiescent nature.

		\section{SED fitting systematics}
		\label{sec:appB}
		
		In order to evaluate our SED modelling method and explore the eventual biases, we produce the simulated data set and fit it using the exact same method that we applied to our observed galaxies. The purpose of using simulations is to analyse eventual observational effects on our SED fitting results. To achieve this goal, we use the functionality available in CIGALE to create mock catalogue of objects for each galaxy for which the physical parameters are known. We build the simulated sample by adopting the best-fit SED model for each fitted QG from our sample. As a result, the procedure gives one artificial model per galaxy. The best SED model per object (with known parameters) is then integrated through the same set of filters as in our observational sample.We then perturb the input fluxes from the best SEDs by adding a randomised noise, following a Gaussian distribution with $\sigma$ corresponding to the observed uncertainty per each photometric band. The SED fitting of galaxies from the mock catalogue is further performed with the exact same choice of physical models and their input parameters as for our real data. This procedure allows to compare the results of the Bayesian analysis provided by CIGALE on the mock catalogue to the input parameters used to build it.
		
\begin{figure}[ht]
			\label{Fig:app1}
			\centering
			\includegraphics [width=5.09cm]{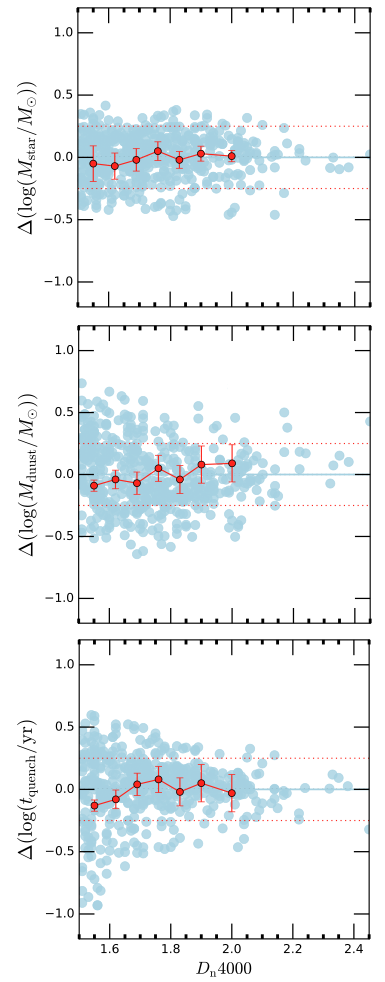}
			\caption{For each panel: results of our mock analysis where we quantify the difference between the input parameters used to build the simulated catalogue in CIGALE, and results of the SED fitting of this mock catalogue. The offset between the simulated and "observed" physical parameters with CIGALE is expressed on y-axis as a $\log$ scale. \texttt{From top to bottom:} stellar mass, dust mass and time since quenching modelled with use of a flexible star formation history. The offset between the "true" and "recovered" values is evaluated as a function of input $D_{\rm n}4000$ from simulated catalogues. We consider that a good constraint on a given parameter is obtained if difference between the input and output approaches zero, which is indicated by the solid blue lines. The trends following the binned means are displayed with red circles with corresponding $1\sigma$ errors, while dotted red lines represent 0.25 dex offset from the zero. Our analysis shows a good overall agreement between the input and output values even in simulated galaxies with higher $D_{\rm n} 4000$.}
		\end{figure}

	The difference (on a log-scale) between the input physical properties and the best output parameters of the simulated catalogue is shown in \hyperref[Fig:app1]{Figure B.1}. All values presented on y-axes are plotted as a function of input $D_{\rm n}4000$, as we want to ensure that our fitting methods does not produce significant systematics as a function of stellar population age. We consider that a strong constraint on a given parameter is obtained if difference between the input and output approaches zero, followed by the small average dispersion. We find that for all the physical quantities analysed in this work, the dispersion of recovered values (input-output) follows the normal distribution, with more than $80\%$ of sources lying within the mean offset of $\pm 0.25$. As expected, slightly dispersed distribution is observed for $t_{\rm quench}$. Nevertheless, despite the known difficulty in fully constraining this parameter (see e.g., \citealt{ciesla21}), the overall trend with $D_{\rm n}4000$ is flat, which assures that our choice of SFH is well-suited for most of our QGs. Therefore, we conclude that our SED fitting procedure properly recovers the true physical properties of our objects and does not introduce any significant systematics.
		
		
		
			\section{Testing the different selections of QGs}
		\label{App:A3}
		
As demonstrated by many studies investigating QGs, all samples that meet the certain quiescent galaxy criteria produce some level of contamination from SF interlopers (see i.e., \citealt{moresco13}, \citealt{siudek18}, \citealt{leja19}, \citealt{bisigello20}, \citealp{zhang22}). Here we test up to which level our results may depend on the method used for selecting the dusty QGs. To accomplish this, we apply some additional, commonly used definitions of quiescence to the main sample of 548 QGs analysed in this work. Namely, we consider QG selection based on UVJ rest-frame colours (e.g., \citealt{muzzin13} and \citealt{schreiber15}), and selection based on more conservative spectroscopic criteria ($D_{\rm n}4000>1.6$, e.g., \citealt{brinchmann04}, \citealt{gallazzi05}). In this way, we quantify the level of potential contaminants, assuming that our alternative selection criteria is providing a "clean" sample of QGs. Normally, this is a very conservative assumption, as it is known that no selection can return a pure sample of QGs. For example, for the full hCOSMOS catalogue, \cite{damjanov18} used the selection proposed by \cite{muzzin13} to separate SF and QGs. They found $\sim10\%$ of objects with $D_{\rm n}4000<1.5$ that would be considered QGs based on UVJ selection. In contrast, for the colour-selected SF galaxies, they discover $13\%$ of objects that would be considered QG based on spectroscopic criteria ($D_{\rm n}4000>1.5$).

Here we perform the similar comparison with the goal to check if selection of dusty QGs introduces significantly larger number of contaminants. For the main UVJ colour selection, we consider \citealt{muzzin13}, who proposed the following criteria to separate star-forming and passively evolving galaxies at $0<z<0.6$:
\begin{equation}
\begin{aligned}
U-V>1.3\\
V-J<1.5 \\
U-V>0.88\times (V-J)+0.69
\end{aligned}
\end{equation}

We also check another popular, redshift invariant UVJ selection proposed in \cite{schreiber15}. It is defined as: 
\begin{equation}
\begin{aligned}
U-V>1.3\\
V-J<1.6 \\
U-V>0.88\times (V-J)+0.49
\end{aligned}
\end{equation}

\begin{figure}[ht]
	\label{Fig:app4}
	\centering
	\includegraphics [width=9.09cm]{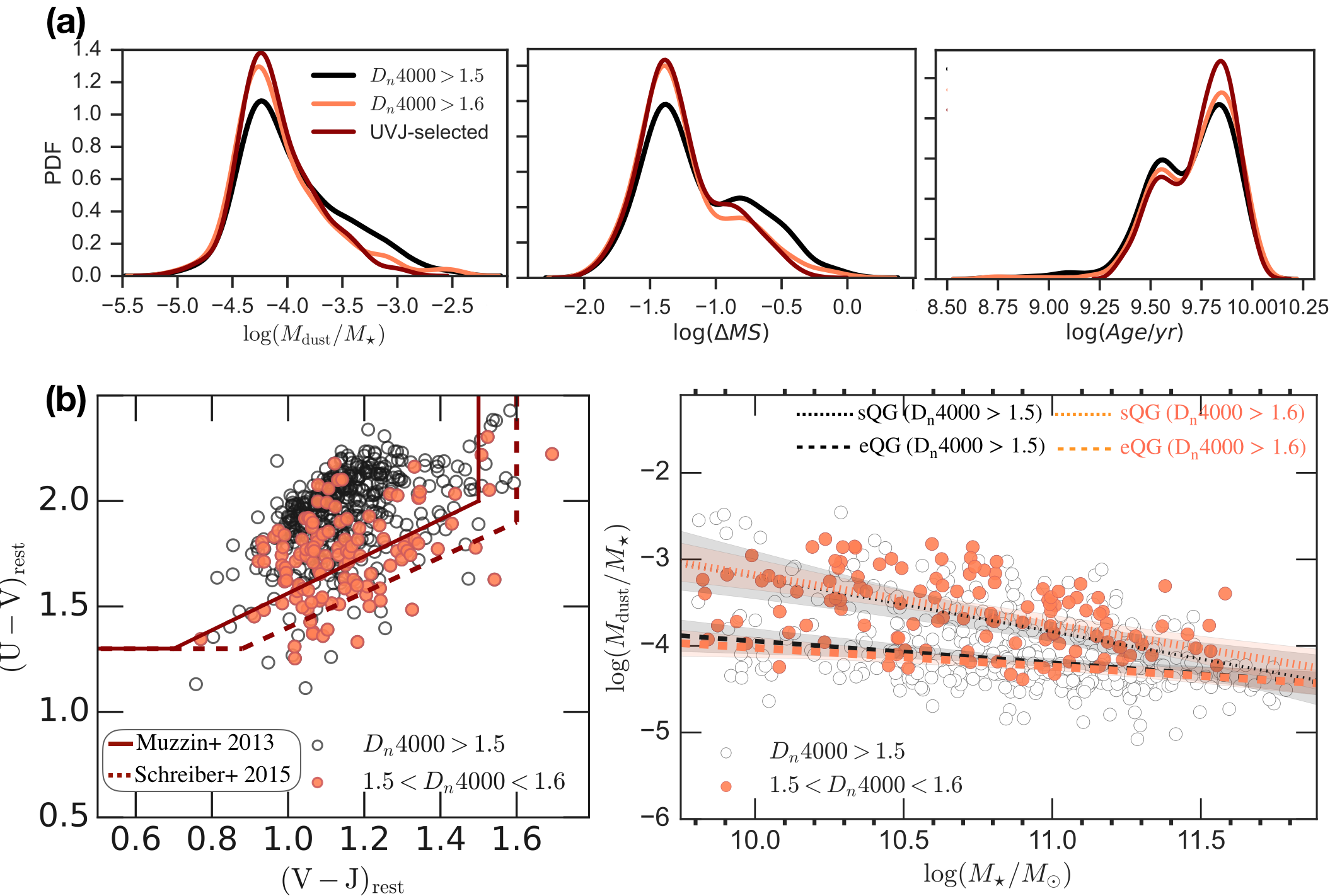}
	\caption{\texttt{Upper panel:} Distributions of $M_{\rm dust}/M_{\star}$, $\Delta_{\rm MS}$ and stellar age for our parent sample (solid black line), against the samples satisfying additional criteria of quiescence: UVJ selection proposed in \cite{muzzin13} (dark red line) and $D_{\rm n}4000>1.6$ (orange line); \texttt{Lower panel:} Positions of 548 dusty QGs from this work in the UVJ rest-frame colour plane. Regions defined with solid and dashed lines correspond to criteria indicated in the legend. Coloured circles represent QGs with $1.5<D_{\rm n}4000<1.6$ and can be considered potential contaminants; \textit{right:} Evolution of specific dust mass as a function of $M_{\star}$. The meaning of symbols is the same as in the left panel. Best linear fits are shown for eQGs and sQGs for two different cases: the main selection that includes 548 QGs (dark grey lines), and selection fulfilling stricter spectroscopic criterion $D_{\rm n}4000>1.6$ (salmon coloured lines). }
\end{figure}

In this analysis, we stay conservative and consider a "contaminant" any outlier at the intersection of our main selection ($D_{\rm n}4000>1.5$) and the alternative one. We then compare the resulting distributions of some key physical parameters (specific dust masses, stellar ages, and the offset from the MS) for such selected QG samples to the distributions from our parent sample presented in the main text. Results of this analysis are shown in \hyperref[tab:AppTab]{Table C.1} and \hyperref[Fig:app4]{Fig.C.1}. 

Regardless of the criteria, it is evident that the vast majority of our 548 dusty QGs reside in the quiescent population region defined by galaxy rest-frame UVJ colours. The fraction of our dusty QGs that satisfy these additional criteria is $83\%$ for the UVJ selection of \cite{muzzin13}, and $94\%$ in the case of the UVJ selection proposed by \cite{schreiber15}. Furthermore, if we impose more strict spectroscopic criterion ($D_{\rm n}4000>1.6$), we would narrow down our parent sample to 420 objects, or $77\%$ of the initial sample. Therefore, the sample  "purity" in all three cases is very high, ranging from $77\%$ to $94\%$ (as the fraction of contaminants varies between $6-23\%$, see \hyperref[tab:AppTab]{Table C.1}). We thus ensure that the selection of IR-detectable "dusty QGs" does not produce a significant impact on the sample purity compared to the samples that have no IR-detection. It is worth noting that the selection by UVJ criteria proposed in \cite{schreiber15} has recently been applied for sampling the "dusty QGs" at $z\sim1.5$ in the GOODS-S field (\citealt{sase23}). Therefore, our parent sample, which almost entirely satisfies that selection, can be used as a benchmark for probing the dust evolution against the stacking samples analysed at higher-$z$'s.

\begin{table}[h]
	\caption{Cross-section between the parent sample of 548 QGs with other selection criteria of QGs}
	\label{tab:AppTab} 
	\centering   
    \scalebox{0.79} {\begin{tabular}{c c c} 
			\hline
			\toprule 
			\\[-13pt]\\
			Quiescent selection criteria&Number of cross-matches&$\%$ of outliers\\[-7pt]\\
		     &with the main sample&\\[-7pt]\\\\
			\hline
		
			       \\[-13pt]\\
			UVJ (\citealp{muzzin13})&453 &$17\%$ \\[-7pt]\\
			
			UVJ (\citealt{schreiber15})&518&$6\%$\\[-7pt]\\
			$D_{\rm n}4000>1.6$&420&$23\%$\\[-7pt]\\
		
			
			\hline    
			\bottomrule               
		\end{tabular}
}
\end{table} 

Upper panel of \hyperref[Fig:app4]{Fig.C.1} shows that, irrespective of applied definitions of quiescence, specific dust masses, stellar ages and $\Delta_{\rm MS}$ follow the same distribution as our main sample of QGs selected as $D_{\rm n}4000>1.5$. Strong agreements in dust-related properties among samples selected in different ways, ensure that the selection of IR-detectable "dusty QGs" does not produce significant impact on the sample purity. As expected, removal of these outliers produces a relative decrease in the number of recently quenched objects from the tail the specific dust mass distribution. However, we check and find that "outliers" have all characteristics of QGs, no detection of $H\alpha$ emission lines, and very prominent negative offset from the MS (median of $\log(\Delta_{\rm MS})=-0.82$). 
As illustrated in the lower right panel of \hyperref[Fig:app4]{Fig.C.1}, our conclusions regarding morphological impact on the evolution of dust-related parameters are entirely preserved even if we change the selection criteria. Therefore, we conclude that the findings and conclusions in this paper are not affected by the definition of quiescence.

	\end{appendix}
	
\end{document}